\documentclass{jfm}
\usepackage{soul}
\usepackage[normalem]{ulem}

\usepackage[utf8]{inputenc}
\usepackage{newtxtext}
\usepackage{amsmath}
\usepackage{newtxmath}
\usepackage{natbib}
\usepackage{hyperref}
\usepackage{color}
     \usepackage{xcolor}
\usepackage{epsfig}
\usepackage{feynmp-auto}
\usepackage{multirow}
\usepackage{tabularx}
\usepackage{array}
\hypersetup{
    colorlinks = true,
    urlcolor   = blue,
    citecolor  = black,
}

\def\bp{\mathbf{p}}

\def\mR{\mathcal{R}}
\def\mI{\mathcal{I}}

\def\mM{\mathcal{M}}
\def\mJ{\mathcal{J}}
\def\mP{\mathcal{P}}

\def\mA{\mathcal{A}}
\def\mB{\mathcal{B}}
\def\mF{\mathcal{F}}
\def\bbR{\mathbb R}

     \usepackage{color}
     \usepackage{xcolor}
\definecolor{g-blue}{rgb}{0.83,0.95,1}
\definecolor{g-yellow}{rgb}{1,1,0.7}
\definecolor{g-green}{rgb}{0.9,1,0.9}
\definecolor{green}{rgb}{0,0.6,0}
\definecolor{cyan}{rgb}{0,0.7,0.7} 
\definecolor{grey}{rgb}{0.4 ,0.4 ,0.4 }
\definecolor{brown}{rgb}{0.6 ,0  ,0.8 }

\def\g-blue#1{\textcolor{g-blue}{#1}}

\usepackage[normalem]{ulem}
 
\title{The Structure of Energy Fluxes in Wave Turbulence}

\author{Giovanni Dematteis and Yuri V. Lvov }

\date{May 2022}

\begin{document}
\maketitle

\abstract We calculate the net energy per unit time exchanged between
two sets of modes in a generic system governed by a three-wave kinetic
equation. Our calculation is based on the property of detailed
  energy conservation of the
  triadic resonant interactions. In a first application to isotropic
systems, we rederive the standard formula for the energy flux as a
particular case for adjacent sets. We then exploit the new formalism
to quantify the level of locality of
the energy transfers in the example of surface capillary waves.
%These results establish a wave turbulence analog of the formalism
%of~\cite{kraichnan1959structure} for energy fluxes in hydrodynamic turbulence.
A second application to anisotropic wave systems expands the currently
available set of tools to investigate magnitude and direction of the
energy fluxes in these systems. We illustrate the use of the formalism
by characterizing the energy pathways in the oceanic internal
wavefield. Our proposed approach, unlike traditional approaches, is
not limited to stationarity, scale-invariance and strict locality.
In addition, we define a number $w$ that quantifies the scale
separation necessary for two sets of modes to be energetically
disconnected, with potential consequences in the
interpretation of wave-turbulence experiments.
The methodology presented here provides a
general, simple and systematic approach to energy fluxes in wave
turbulence.

\section{Introduction}\label{sec:1}

Wave turbulence has a six-decade-long successful record in describing
inter-scale energy transfers in nonlinear wave media in geophysics --
internal inertia-gravity waves \citep{olbers1976nonlinear,LT}, surface
gravity waves \citep{hasselmann62,zakharov1967energy} and capillary
waves \citep{zakharov1967weak}, Rossby waves
\citep{zakharov1988canonical}, inertial waves \citep{galtier2003weak}
--, astrophysics -- e.g. plasma
\citep{sagdeev1969nonlinear,zakharov1972collapse}) --, solid state
physics \citep{ziman2001electrons}, acoustic waves
\citep{Zakharov:70}, vibrating plates \citep{during2006weak} and
Bose-Einstein condensates \citep{NazBook}.

In addition to a close formal similarity to hydrodynamic turbulence,
the large theoretical relevance of wave turbulence is related to the
derivation of nonequilibrium cascade states known as the
Kolmogorov-Zakharov (KZ) solutions \citep{VLvov1992}. Unlike the
``dimensional'' Kolmogorov spectrum of 3D turbulence, the KZ spectra
are analytical solutions of the equation that represents the
  main object of wave turbulence theory, namely the wave kinetic
  equation (WKE). The WKE describes the time evolution of the spectral energy density
  due to the nonlinear resonant energy transfers between different
  wave modes.

Non-zero inter-scale energy fluxes are a fundamental feature of wave
turbulence that is still far from being fully understood -- see
e.g. the recent works \cite{hrabski2022properties} and
\cite{dematteis_lvov_2021}.  In geophysical applications, the study of
wave turbulence fluxes dates back to the early '80s for internal waves
\citep{MM81,M86} and surface gravity waves
\citep{hasselmann1981symmetrical}. Those early studies relied mainly
on diffusive approximations of the collision operator, the r.h.s. of
the WKE that describes the irreversible modal energy transfers due to
wave-wave interactions. Subsequent improvements of the approximations
to flux computations led to important theoretical and numerical tools
that are used to this day. For the surface gravity wave problem, the
numerical schemes currently employed in the WAM global model of wave
forecasting
\citep{hasselmann1985computations,resio1991numerical,komen1996dynamics,janssen2004interaction}
use approximations of the main resonant wave quartets that are
responsible for the direct and inverse cascade of energy and wave
action through the wave spectrum. This allows for accurate predictions
of the global sea states, explaining for instance the formation of the
large oceanic swells from an inverse cascade process toward the long
waves.  In the ocean interior, the inter-scale fluxes in the oceanic
internal wavefield due to resonant wave triads are the backbone of the
\cite{gregg1989scaling}-\cite{henyey1991scaling}-\cite{polzin1995finescale}
finescale parameterization of oceanic mixing and dissipation, a
fundamental component of the global models of ocean circulation
\citep{mackinnon2017climate,whalen2020internal,polzin2009abyssal,musgrave2022lifecycle}. The
scaling of this phenomenological parameterization is based on the
``induced diffusion'' approximation of the WKE of internal
waves~\citep{McComas1977}. In the finescale parameterization
framework, a downscale flux in the internal waves is associated to the
production of mixing and dissipation by the turbulence that is
generated when the internal waves overturn and break due to
hydrodynamic instabilities. This mixing allows for bottom dense
  water to slowly upwell toward the surface at low
  latitudes, with major consequences on
the meridional overturning circulation in the
ocean~\citep{thorpe2005turbulent,garabato2022ocean}. Both of these
notable examples, oceanic surface and internal waves,
require understanding of the inter-scale
fluxes being transferred through a random bath of resonantly
interacting waves. This understanding is important not only for the
quantification of the wavefield itself, but also for the paramount
implications of the coupling of these systems with other components of
the climate system.

The approximation schemes mentioned above make use of uncontrolled,
often empirical approximations. From a theoretical perspective, the
computation of energy fluxes from the collision operator of the WKE is
elusive, since the collision operator itself is vanishing in a
stationary state. For {the} KZ spectra, as explained in
\cite{VLvov1992} and in Sec.~\ref{sec:3.1.1} below, there is an
indeterminate expression of the type of $0/0$ which requires
regularization (using de L'H\^{o}pital's rule). The flux is thus given
by the coefficient of the next-order term in a Taylor-series expansion
of the collision operator, centered in the KZ exponent
\citep{VLvov1992}. However, for non-KZ stationary states, which are
relevant solutions e.g. in anisotropic systems like Rossby waves
\citep{NazBook} and internal waves \citep{iwthLPTN}, in general it is
not clear how to calculate the flux from the collision operator.
Moreover, some of the early quantifications of energy transfers failed
to notice the key difference between the energy density time increment
and the actual energy flux. The idea can be explained with the help of
a 1D example.  Let $\dot e_p$ be the energy density rate of change,
and $F_p$ the energy flux, where $p$ is the scalar wavenumber
variable. In general the energy balance for an infinitesimal interval
$[p,p+dp]$ reads: $\dot e_p dp=F_p-F_{p+dp}$. A slightly positive
$\dot e_p$ could correspond to a negative or positive flux alike, as
long as $F_p$ is a decreasing function of $p$. When $\dot e_p$ is
vanishing instead -- which defines stationary conditions -- the flux
is constant in $p$, but its value cannot be calculatd from $\dot e_p$
alone.  Thus, the sign of $\dot e_p$ is quite unrelated to the
direction and magnitude of $F_p$!  This objection was put forward in
\cite{holloway1980oceanic}, arguing that close to a stationary
  state the small value of $\dot e_p$ has nothing to do with the
  timescale of the energy pathways, or more precisely with the
  ``residence time'' of energy in the wavefield. This timescale is
  dictated by the magnitude of $F_p$. Even
when the difference between the two quantities has been treated
correctly, much more emphasis has been given in the literature to the
evaluation of the rate ($\dot e_p$ in the intuitive example) rather
than to the actual flux ($F_p$).  Finally, another remarkable
theoretical need is the generalization of the theory of the fluxes of
conserved quantities to wave systems that are not self similar, since
the bulk of the theory was
mainly developed for scale-invariant
spectra~\citep{VLvov1992,NazBook}.

To summarize, the existing body of
literature is focused on calculating energy fluxes in stationary
isotropic scale-invariant wave turbulence systems. Yet, the kinetic
equation contains a lot of information about wave-wave interactions
not currently utilized. Here, we propose a way to calculate energy
fluxes that is free of these limitations.

In this work, we focus on the study of {\it three-wave collision
  operators} and tackle the problem of quantifying the associated
energy transfer between two generic disjoint {\it control volumes} in
Fourier space. We introduce a logical operator, namely the {\it
  characteristic interaction weight}; this weight allows us to extract
the flux between the two control volumes from the collision operator,
by singling out those triads of wavenumbers that imply a direct
energetic link between the control volumes themselves. The definition
of the {\it characteristic interaction weight} is based on a
fundamental symmetry of the three-wave collision operators, namely the
{\it detailed energy conservation} property \citep{kraichnan1959structure}.  As a result, any non-vanishing energy fluxes, even
in a stationary state, can be calculated by integration of a
well-defined non-vanishing function. We call this function the {\it
  transfer integral} of the problem. Note that these calculations are
exact: they do not employ any approximation other than the assumption
of validity of the wave kinetic equation. Moreover, self-similarity is
not required. Our results establish a formal wave-turbulence parallel
to the \cite{kraichnan1959structure} computation of energy fluxes for
hydrodynamic turbulence at high Reynolds numbers, versions of which
have been used for different models of turbulence (e.g., see
\cite{kraichnan1975remarks}, \cite{rose1978fully} and \cite{eyink1994energy}).

In our approach, we postulate a governing wave kinetic equation with
an inertial range of scales. In support to this {\it kinetic
  assumption}, we appeal to the current fervent research toward a
rigorous justification of the WKE from the deterministic equations of
motion
\citep{choi2004probability,NazBook,lukkarinen2011weakly,eyink2012kinetic,chibbaro20184,onorato2020straightforward,
  buckmaster2021onset,rosenzweig2022uniqueness,deng2021full,deng2021propagation,banks2022direct}.

The manuscript is organized as follows. In the remainder of
Sec.~\ref{sec:1}, we set up the stage by introducing the WKE and
its relevant properties. Sec.~\ref{sec:2}
contains our {\it Main Statement} in the form of a formula for the computation of energy transfers
between two generic control volumes in spectral space. Its
application to isotropic systems is treated in Sec.~\ref{sec:3}, where
the standard flux formula of isotropic wave turbulence is derived
rigorously as a particular case of the {\it Main Statement} for adjacent control volumes, and the
concept of transfer integral is defined. In Sec.~\ref{sec:4} we
illustrate the results for the surface-capillary-wave example,
including a detailed quantification of the locality properties of the
system. Sec.~\ref{sec:5} is devoted to the application to anisotropic
systems, followed by a practical illustration for the internal wave
problem in Sec.~\ref{sec:6}. In Sec.~\ref{sec:7} we exploit the
transfer-integral formulas previously derived to calculate the
convergence conditions for the energy flux and to define a number $w$ quantifying the level of locality of the energy transfer. We discuss and summarize our results in
Sec.~\ref{sec:8}.

\subsection{Wave kinetic equation}
We start from the Wave Kinetic Equation of a system with three-wave resonant interactions~\citep{VLvov1992,NazBook},
    \begin{equation}\label{eq:1}
    \begin{aligned}
        &\frac{\partial n_\bp}{\partial t} = \int_{\bbR^d\times \bbR^d} d\bp_1 d\bp_2 \mJ(\bp;\bp_1,\bp_2)\,,\quad\mJ(\bp;\bp_1,\bp_2) = \mR^0_{12} - \mR^1_{02} - \mR^2_{01}\,,\\
        &\text{where}\quad \mR^0_{12} = 4\pi |V^0_{12}|^2 f^0_{12}\delta(\bp^0_{12})\delta(\omega^0_{12})\,,\quad f^0_{12} = n_1n_2-n_\bp(n_1+n_2)\,,
        \end{aligned}
    \end{equation}
and $\bp^0_{12}=\bp-\bp_1-\bp_2,
\omega^0_{12}=\omega_\bp-\omega_1-\omega_2$.  The variable $n_\bp$ is
the $d-$dimensional wave-action spectral density at wavenumber
$\bp\in\bbR$.  For simplicity we denote $p_i$ by its index $i$ in
subscripts and superscripts, and the wavenumber variable $\bp$ by
index $0$.  Action can be viewed as the ``number'' of waves with a
given wavenumber. The function $\omega_\bp$ is the linear dispersion
relation of the system, taking the positive branch by
convention. Consequently, wave action multiplied by frequency
$\omega_\bp n_\bp$ is the quadratic spectral energy density. Note that
wavenumbers are vectors in $\bbR^d$, while frequencies are always
positive scalars. The factor $V^0_{12}$ is the interaction matrix
element (or scattering cross-section) describing the transfer of wave
action among the members of a triad composed of three wavenumbers
$\bp,\bp_1,\bp_2$. $V^0_{12}$ is invariant under permutation of the
indeces $1$ and $2$, and therefore so is $\mR^0_{12}$. We refer to
$\mJ(\bp;\bp_1,\bp_2)$ as the interaction kernel (or collision
integrand) associated with the given WKE. The r.h.s. of~\eqref{eq:1}
is then called the collision integral, a quadratic functional in the
action density $n_\bp$. The collision integral captures the
irreversible transfers of action between different modes as the
outcome of nonlinear interactions between triads of wavenumbers {\it
  in resonance} with each other.
    
\subsection{Resonant manifold}
Let the dispersion relation of the system be of the form $\omega_\bp =
\omega(|p_1|,...,|p_d|)$, positive definite, monotonic in each
component, and such that it allows for non-trivial solution of the
three resonant conditions
    \begin{equation}\label{eq:2}
        {\rm(I)}:\left\{\begin{aligned}&\bp = \bp_1+\bp_2\,,\\&\omega_\bp=\omega_1+\omega_2\,\end{aligned}\right.,\qquad
        {\rm(II)}:\left\{\begin{aligned}&\bp_1 = \bp+\bp_2\,,\\&\omega_1=\omega_\bp+\omega_2\,\end{aligned}\right.,\qquad
        {\rm(III)}:\left\{\begin{aligned}&\bp_2 = \bp+\bp_1\,,\\&\omega_2=\omega_\bp+\omega_1\,\end{aligned}\right..
    \end{equation}
For instance, limited to power-law dispersion relations
$\omega(\bp)\propto|\bp|^\alpha$, the condition $\alpha>1$ is
necessary and sufficient for the existence of solutions to
Eqs.~\eqref{eq:2}~\citep{VLvov1992}.

Note that the invariance upon permutation of the indices $1,2$ in
$|V^0_{12}|$ and $f^0_{12}$ allows to express the interaction kernel
as $\mJ(\bp;\bp_1,\bp_2) = \mR^0_{12} - 2\mR^1_{02}$. This, in turn
allows in a completely general way only have to deal with resonance
types I (i.e. the sum interactions), and II (i.e. the difference
interactions). To simplify the notation in the following, let us
denote by $\mJ^{(l)}(\bp,\bp_1,\bp_2)$, with $l=$ I, II, III, the
three terms $\mR^0_{12} ,-\mR^1_{02},-\mR^2_{01}$, respectively.

In general, the WKE can then be reduced to
    \begin{equation}\label{eq:4}
    \frac{\partial n_\bp}{\partial t} = \sum_l \int_{\Omega_l} d\Omega_l J^{(l)}\,, \quad\text{with } l={\rm I,II,III}\,,
    \end{equation}
    where $J^{(l)}$ is the result of analytical integration of the
    $(d+1)$ independent delta functions and $\Omega_l$ is a
    $(d-1)$-dimensional parameterization of the respective branch of
    the resonant manifold. Note that each of the resonant conditions
in~\eqref{eq:2} can have multiple independent solutions (c.f. Sec.~\ref{sec:6}), in which case in Eq.~\eqref{eq:4}
a summation over the independent solutions of each branch is implicit. Once the WKE collision integral is
    suitably expressed in the form~\eqref{eq:4}, integration over the
    remaining $d-1$ degrees of freedom can be performed. The integration can be
        performed either analytically or numerically depending on the particular situation.

\subsection{Detailed energy conservation}

We end this introductive section highlighting a fundamental property
of the interaction kernel. The three-wave resonant interactions in the
collision integral~\eqref{eq:1} satisfy {\it detailed energy
  conservation}~\citep{onsager1949statistical,kraichnan1959structure,hasselmann1966feynman,rose1978fully,eyink1994energy}:
\begin{quote}
  {\bf Property: Detailed energy conservation.} \it We define the
  quantity
\begin{equation}{\cal Z}(\bp_{a},\bp_b,\bp_c)=
  \omega_a\mJ(\bp_{a};\bp_b,\bp_c) + \omega_b\mJ(\bp_{b};\bp_a,\bp_c) + \omega_c\mJ(\bp_{c};\bp_a,\bp_b).\nonumber\end{equation}
Then, for any given triad of
wavenumbers $\bp_a, \bp_b, \bp_c$
\begin{equation}\label{eq:DC}
 {\cal Z}(\bp_{a},\bp_b,\bp_c)=0.
\end{equation}
\end{quote}

A proof is provided in {\it Appendix}~\ref{app:A}.
We note that the equality holds in the sense of distributions, since ${\cal Z}(\bp_{a},\bp_b,\bp_c)$ contains delta functions.

The physical meaning of ${\cal Z}(\bp_{a},\bp_b,\bp_c)$ is the amount
 of energy generated during the triadic interactions of three wave
 numbers. This quantity is zero due to energy conservation, as ensured
 by  the frequency delta functions.
      We note that this property holds for triads of
     wavenumbers on the resonant manifold (\ref{eq:2}), as well
     as for triads of wavenumbers {\it off} the resonant manifold.

     \section{Energy transfer between two disjoint sets
of wavenumbers}\label{sec:2}

Equation~\eqref{eq:1} is derived under the assumption that the
quadratic energy is a good approximation of the total energy of the
system. The quadratic energy density $\omega_\bp n_\bp$ is exactly
preserved by the time evolution of~\eqref{eq:1}, representing what is
sometimes referred to as an adiabatic invariant (see e.g. section
8.5.1 in~\cite{NazBook}). Mathematically, this property is enforced by
the frequency delta function in the collision integral, which can be
interpreted as the condition of energy conservation in the triadic
interactions. This is captured by the property of detailed energy conservation~\eqref{eq:DC}.

From now on we will refer to $e_\bp=\omega_\bp n_\bp$ simply as the
spectral energy density. After multiplying equation~\eqref{eq:1} by
$\omega_\bp$, the r.h.s. contains the energy transfers between
wavenumber $\bp$ and all possible pairs of wavenumbers $\bp_1$ and
$\bp_2$ which interact resonantly with $\bp$.

Let us consider $A\subset \bbR^d$ and $B\subset \bbR^d$, with
$A\cap B=\emptyset$, two disjoint closed subsets of the
$d$-dimensional Fourier space.  For a given specification of the
action spectrum $n_\bp$, we wish to quantify how much power (energy
per unit of time) is transferred instantaneously from set $A$ to set
$B$. The following statement holds:

\begin{quote}\it
{\bf Main Statement: } The net power transferred instantaneously from set A to set B  under the governing resonant
dynamics of Eq.~\eqref{eq:1} is given by
\begin{equation}\label{eq:5}
    \mP_{A\rightarrow B} = -\int_A d\bp \omega_\bp \int_{\bbR^d\times \bbR^d} d\bp_1 d\bp_2 \sum_l\chi_{B}^{(l)}(\bp_1,\bp_2) \mJ^{(l)}(\bp,\bp_1,\bp_2)\,,
\end{equation}
with $l={\rm I,II,III}$, where $\chi_{B}^{(l)}(\bp_1,\bp_2)$ is a characteristic interaction weight defined in Table~\ref{tab:1}.

\begin{table}
\begin{center}
\begin{tabular}{c||c|c|c}
	&$l={\rm I}$ & $l={\rm II}$  & $l={\rm III}$ \\
\hline
	$\chi_{B}^{\rm(l)}(\bp_1,\bp_2)$= 
&  $ \displaystyle    \left\{\begin{aligned}&1\qquad\quad\quad\text{if } \bp_1\in B, \bp_2\in B\\
    &\frac{\omega_1}{\omega_1+\omega_2}\quad\text{if } \bp_1\in B, \bp_2\notin B\\
    &\frac{\omega_2}{\omega_1+\omega_2}\quad\text{if } \bp_1\notin B, \bp_2\in B\\
    &0\qquad\quad\quad\text{otherwise}
    \end{aligned}\right. $
& $ \displaystyle     \left\{\begin{aligned}&1\quad\text{if } \bp_1\in B\\
    &0\quad\text{if } \bp_1\notin B
    \end{aligned}\right. $
& $  \displaystyle   \left\{\begin{aligned}&1\quad\text{if } \bp_2\in B\\
    &0\quad\text{if } \bp_2\notin B
    \end{aligned}\right. $
\end{tabular}
\end{center}
\caption{Specification of the interaction weights in Eq.~\eqref{eq:5}.\label{tab:1}}
\end{table}
\end{quote}

{\noindent{\bf Sketch of the proof.}}

The structure of the collision
integral can be interpreted as follows. Given any two wavenumbers
$\bp_1$ and $\bp_2$ in resonance of type $l$ with $\bp$, the
interaction kernel $\mJ^{(l)}(\bp,\bp_1,\bp_2)$ quantifies how much
wave-action density (per unit time) is being transferred
instantaneously to $\bp$ by the three-wave interaction between the
wavenumbers $\bp$, $\bp_1$ and $\bp_2$. When the term is positive,
contributing to an increment of $n_\bp$, wavenumber $\bp$ is generated
as an output of the interaction. When the term is negative,
contributing to a decrement of $n_\bp$, wavenumber $\bp$ is absorbed
as an input of the interaction. The type of three-wave interaction
(coded by $l$) and the sign of the contribution are enough information
to ``build'' the directed energy diagram associated to the
triad. Then, integrating over all possible combinations of $\bp_1$ and
$\bp_2$ provides the net action increment per unit time for mode
$\bp$, i.e. the l.h.s. $\dot n_\bp$. Multiplying the contribution by
$\omega_\bp$ allows us to quantify the net energy increment per unit
time for mode $\bp$.

%The idea of the proof is to consider all possible variations of
%solutions of~\eqref{eq:2}, i.e. all points on the resonant manifold.
Thus, a triad of wavenumbers $\bp,\bp_1,\bp_2$ on the resonant manifold
leads to an istantaneous change of $n_\bp$ to
$n_\bp+\dot n_\bp|_{012} dt$, where we define
$\dot n_\bp|_{012}:=\mathcal J(\bp;\bp_1,\bp_2)$ (index $0$ is used
here to denote wavenumber $\bp$).  This increases the energy at $\bp$
by a quantity $\dot e_\bp|_{012}dt=\omega_\bp \dot n_\bp|_{012} dt$.
The property of detailed energy conservation~\eqref{eq:DC} can now be
written equivalently as
\begin{equation}\label{eq:seethelight}
\begin{aligned}
	\omega_\bp\dot n_\bp|_{012} + \omega_1\dot n_1|_{012} + \omega_2\dot n_2|_{012} &= 0\,,\quad \text{or}\\
	\dot e_\bp|_{012} + \dot e_1|_{012} + \dot e_2|_{012} &= 0\,.
\end{aligned}
\end{equation}

The reasoning behind the result~\eqref{eq:5} is the following. First, we express $\dot e_1|_{012}$ and $\dot e_2|_{012}$ as a function of $\dot e_0|_{012}$, quantifying how much of the energy transferred to wavenumber $\bp$ comes from $\bp_1$ and how much from $\bp_2$. Secondly, we must consider all possible cases of whether $\bp_1$ and $\bp_2$ are or not in set $B$ to quantify the energy transferred from set $B$ to a generic point $\bp\in A$. The interaction weights $\chi_{B}^{(l)}(\bp_1,\bp_2)$ appear naturally as a result of this calculation. Third, an outer integration over all points $\bp\in A$ yields the total energy transferred from set $B$ to set $A$ per unit time, i.e. an istantaneous power. The key to the proof, found in {\it Appendix}~\ref{app:A}, is the detailed energy conservation property~\eqref{eq:DC}.

%the intrinsic symmetry of the kinetic equation

\section{Isotropic systems}\label{sec:3}
\subsection{Overview on the theory of energy fluxes}\label{sec:3.1.1}
 We start here by revisiting the classical arguments for the spectral
 energy fluxes in wave turbulence. These arguments appear in
 \cite{VLvov1992}. Here we revisit these arguments to prepare the soil
 for additional insights into spectral energy transfers which are
 obtained by using our formalism.  Our first application of
formula~\eqref{eq:5} is to isotropic scale-invariant
 systems.  We assume a power-law dispersion relation allowing for
 three-wave resonant interactions and scale-invariant matrix elements
 with homogeneity exponent $m$:
\begin{equation}\label{eq:disp}
	\omega_p = \kappa p^\alpha\,,\;\; \alpha>1\,,\qquad \left|V_{12}^0\right|^2=V_0^2 p^{2m} f\left(\frac{p_1}{p},\frac{p_2}{p}\right)\,,
\end{equation}
allowing us to look for general solutions to~\eqref{eq:1} of the form
\begin{equation}
	n_p=A p^{-s}\,.
\end{equation}
Let us start by reviewing some classical results for the energy fluxes
in such systems, summarized in Chapter 3 of~\cite{VLvov1992}. In
direct analogy with the local energy cascades in isotropic turbulence,
it is assumed that the interactions are sufficiently local in Fourier
space
\citep{kolmogorov1941local,kraichnan1959structure,rose1978fully,eyink2005locality},
so that one can assume a differential continuity equation for the 1D
spectral energy density
\begin{equation}\label{eq:9}
	\frac{\partial e_p}{\partial t} = \pi (2p)^{(d-1)}\omega_p I_p = -\frac{\partial F}{\partial p}\,,
\end{equation}
where the r.h.s. of the WKE is interpreted as minus the divergence of
a flux $F$. Here, $I_p$ is the collision integral, multiplied by the
area of the $d$-dimensional sphere. Supposed there are no energy
sources or sinks in an {\it inertial range} $[\epsilon, M]$, taking
$\epsilon\to0$ and $M\to\infty$, solving for the flux $F$ one obtains
\begin{equation}\label{eq:11}
	F(p) = -\pi\int_0^p dp'\; (2p')^{(d-1)} \omega_p' I_{p'} \,.
\end{equation}
Interpreting the collision integral as the divergence of a pointwise
flux subtends the intuition that energy transfers happen locally in
Fourier space. The underlying reasoning involves the following
steps. Assume a partition of Fourier space into small boxes of width
$\Delta p$. Assume that the time variation of the energy contained in
the box between $p$ and $p+\Delta p$, say $\dot e_{[p,p+\Delta p]}$ is
only due to the energy exchanges with its two adjacent boxes. Call
$F_p$ the net power exchanged at $p$ and $F_{p+\Delta p}$ the net
power exchanged at $p+\Delta p$. Express energy conservation for the
box under consideration as $\dot e_{[p,p+\Delta p]} = F_p -
F_{p+\Delta p}\,.$ Now, take $\Delta_p\to0$, and obtain~\eqref{eq:9}
by standard transition to a continuum representation. In turbulence,
the conditions on how fast the correlations have to decay for the
transfers to be sufficiently local are studied
in~\cite{kraichnan1959structure} and ~\cite{eyink1994energy}. In wave turbulence,
a transposition of the same arguments leads to the statement that if
the collision integral is convergent, then the interactions are
sufficiently local for the differential conservation
picture~\eqref{eq:9} to hold. For this reason, the convergence
conditions for the collision integral in Eq.~\eqref{eq:1} are named
the {\it locality conditions} \citep{VLvov1992}.  However, we are also
not aware of a rigorous proof of this fact.

When locality holds, the expression for the instantaneous energy
flux~\eqref{eq:11} is valid in general, both in stationary and
nonstationary conditions. The wave turbulence theory of
scale-invariant spectra~\citep{VLvov1992} focusses on the stationary
solutions to~\eqref{eq:1}. These can be equilibrium ($F=0$) or
nonequilibrium ($F=\rm const$) solutions, i.e. the Rayleigh-Jeans and
the Kolmogorov-Zakharov (KZ) solutions, respectively. The KZ spectrum
can be obtained dimensionally or via the Zakharov-Kraichnan conformal
transformations~\citep{zakharov1967weak,zakharov1972collapse} and we
have

\begin{equation}\label{eq:13}
	n_p^{RJ} =A p^{-\alpha}\,,\qquad\qquad n_p^{\rm KZ}=A p^{-s_0}\,,\;\text{with}\quad s_0=m+d\,.
\end{equation}
A paradox (only apparent) has to be solved: a constant flux $F\neq0$
must result from integrating a vanishing integrand in
Eq.~\eqref{eq:11}! It is convenient to switch to $\omega$ space using
the dispersion relation as the change of variables, by defining
\begin{equation}
	I_\omega=\pi (2p)^{d-1}
        \left(\frac{d\omega}{dp}\right)^{-1}I_p\,,\quad \text{so that
        }\quad I_\omega = \omega^{\sigma-2}(V_0A)^2 \mI(s)\,,
\end{equation}
where $\mI(s)$ is a nondimensional integral that vanishes in the
stationary states, \ and $\sigma = 2(m+d-s)/\alpha$.  Now,
Eq.~\eqref{eq:11} reads

\begin{equation}\label{eq:14}
	F(\omega) =-  \int_0^\omega d\omega'\; \omega' I_{\omega'} = -\omega^\sigma (V_0A)^2 \frac{\mI(s)}{\sigma}\,.
\end{equation}
At the KZ solution~\eqref{eq:13}, we have $\sigma=0$, and therefore an
indeterminate form $0/0$. This indeterminate form is then regularized
by Taylor-expanding $\mI(s)$ to first order, or equivalently by using de L'H\^{o}pital's rule. We thus obtain
\begin{equation}\label{eq:16}
	F = - (V_0A)^2 \left.\frac{d\mI}{ds}\right|_{s=s_0}\,,
\end{equation}
where the locality conditions ensure that
$\left.{d\mI}/{ds}\right|_{s=s_0}$ is finite, with the property that
the flux is positive if $s_0>\alpha$, i.e. the KZ spectrum is steeper
than the equilibrium spectrum. The solution does not exist if
$s_0<\alpha$.  Moreover, note that $F$ is independent of $\omega$,
consistently with stationarity and corresponding to a constant
downscale energy flux in the wave turbulence inertial range.
 
\subsection{Application of the {\it Main Statement}~\eqref{eq:5} to isotropic systems}
Using integration variables in $\omega$-space, in isotropic conditions
Eq.~\eqref{eq:1} simplifies to
%    \begin{equation}\label{eq:17}
 %   \begin{aligned}
%        &\frac{\partial n_p}{\partial t} = \frac{v_p}{p^{d-1}} \int\int_{0}^\infty d\omega_1 d\omega_2 \;  J(\omega_p;\omega_1,\omega_2) \,,\quad J(\omega_p;\omega_1,\omega_2) = R^0_{12} - 2 R^1_{02} \,, \\
 %       &\text{where}\quad R^0_{12} = 4\pi  \sigma^{3(1-d)/\alpha} \frac{(\omega\omega_1\omega_2)^{\frac{d-1}{\alpha}}}{v_p v_1v_2} \frac{|V^0_{12}|^2 f^0_{12}}{\Delta_d} \delta\left(\omega^0_{12}\right)\,.
%        \end{aligned}
%    \end{equation}
\begin{equation}\label{eq:17}
        \begin{aligned}
        &\frac{\partial n_p}{\partial t} = \frac{v_p}{p^{d-1}}\int_{0}^\infty d\omega_1 \left( J^{({\rm I})}(\omega_p;\omega_1,\omega_p-\omega_1)+2 J^{({\rm II})}(\omega_p;\omega_1,\omega_1-\omega_p)\right) \,, \\
        &\text{where}\qquad J^{({\rm I})}(\omega_p;\omega_1,\omega_2) = R^0_{12}\,,\quad J^{({\rm II})}(\omega_p;\omega_1,\omega_2) = - R^1_{02}\,, \\
&\text{and}\qquad\quad R^0_{12}  = 4\pi  \kappa^{3(1-d)/\alpha} \frac{(\omega\omega_1\omega_2)^{\frac{d-1}{\alpha}}}{v_p v_1v_2} \frac{|V^0_{12}|^2 f^0_{12}}{\Delta_d}\,.
        \end{aligned}
    \end{equation}
    We have used the notation $v_p = \partial \omega/\partial p $ and
    $\Delta_d$ is defined by angle integration of the wavenumber delta
    function given space isotropy, with the dimensions of a wavenumber
    to the $d$-th power. We assume a scale invariant solution
\begin{equation}\label{eq:e3.10}
	n_p = A \omega_p^{-x}\,.
\end{equation}
Using these variables, saying that the interaction kernel $J(\omega_p;\omega_1,\omega_2)$ has a homogeneity exponent of $\gamma_0-2x$, the KZ solution has exponent $x=(\gamma_0+3)/2$, and the RJ solution has an exponent $x=1$.
\begin{figure}
\begin{centering}
\includegraphics[width=.5\linewidth]{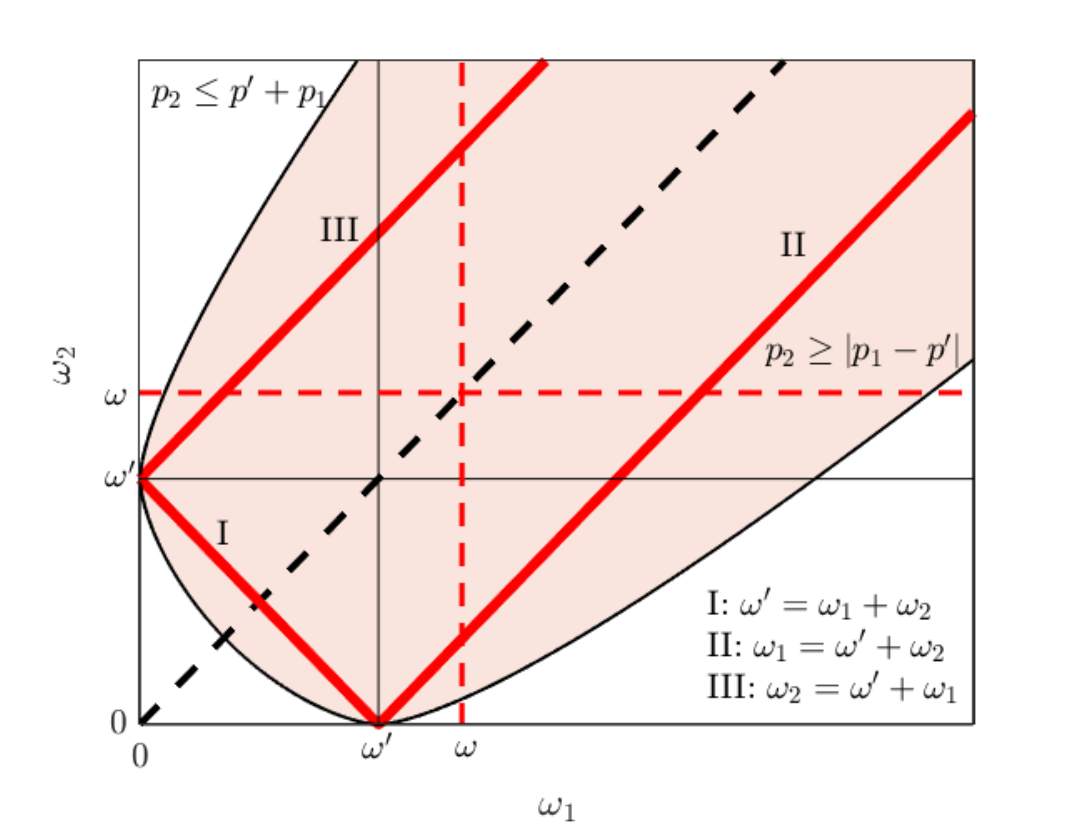}%
\includegraphics[width=.5\linewidth]{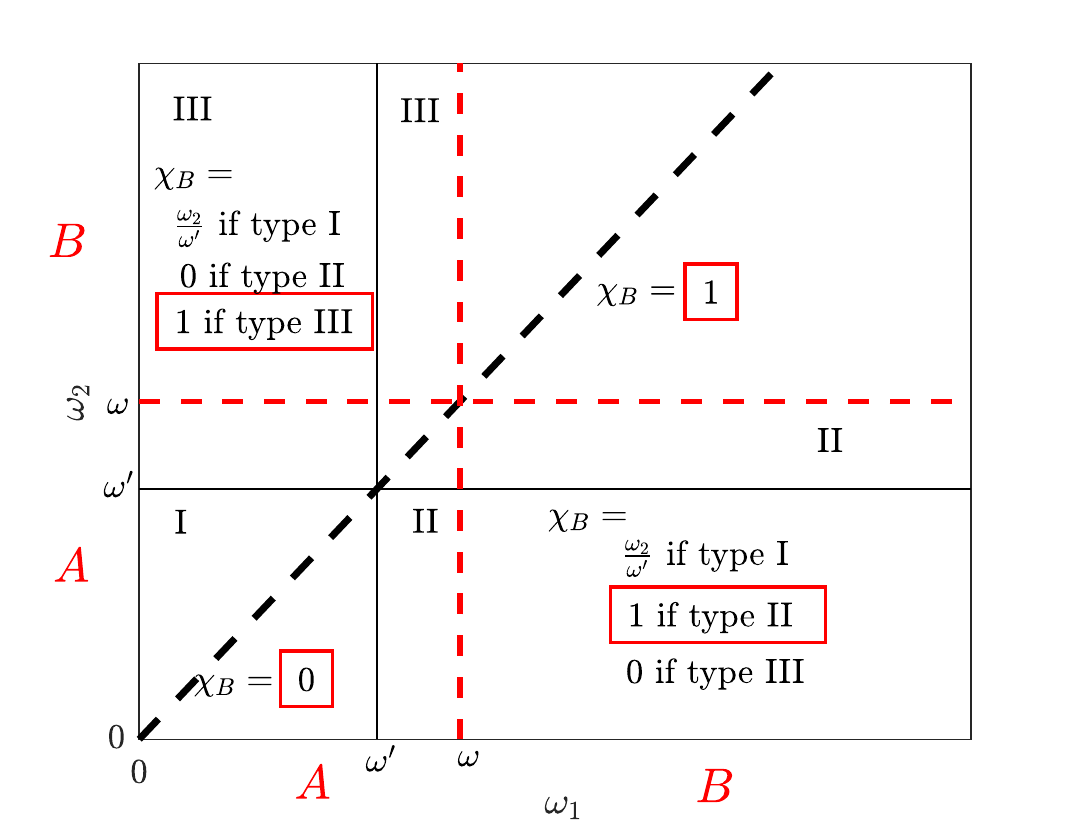}
  \caption{Left: Representation of the resonant manifold in the
    $\omega_1-\omega_2$ space for triads involving wavenumbers
    $\omega'$, $\omega_1$, $\omega_2$. Here $\omega$ demarks the
    separation between sets $A$ and $B$ (dashed red lines). The three
    resonant branches are represented as the three solid red lines
    labeled by I, II, III. The shaded area denotes the region
    satisfying the wavenumber delta function condition, for a value
    $\alpha>1$ (if $\alpha<1$, this area becomes disjoint from the
    frequency condition lines, and there are no resonances).  Right:
    The values of the characteristic interaction weight
    $\chi_{B\rightarrow \omega'}$ in the different regions, as given
    by the {\it Main Statement}~\eqref{eq:5}. Highlighted by a red rectangle are the
    relevant cases to each region, due to the interaction type present
    in that region. Exploiting the symmetry with respect to the main
    diagonal and considering only $\omega_2\le \omega_1$ (below the
    dashed black line), it is clear that $\chi_{B\rightarrow
      \omega'}=\Theta(\omega_1-\omega)$.}
 \label{fig:3}
\end{centering}
\end{figure}

By formula~\eqref{eq:5}, given $A$ and $B$ two disjoint closed subsets
of Fourier space (spanned by $\omega\in\bbR^+$), the instantaneous
power delivered from $A$ to $B$ amounts to
\begin{equation}\label{eq:19}
\begin{aligned}
    \mP_{A\rightarrow B} & = -2^{d-1}\pi \int_A d\omega' \;\omega' \int_0^\infty d\omega_1   \left(\chi_{B}^{({\rm I})}(\omega_1) J^{({\rm I})}(\omega';\omega_1,\omega'-\omega_1)\right.\\
&\qquad\qquad\qquad\qquad\qquad\left.+2 \chi_{B}^{({\rm II})}(\omega_1) J^{({\rm II})}(\omega';\omega_1,\omega_1-\omega')\right) \,,
\end{aligned}
\end{equation}
where the dependence on $\omega_2$ in the interaction weigth is
implicitly constrained by $\omega_2=\omega'-\omega_1$ in the first
line and by $\omega_2=\omega_1-\omega'$ in the second line.  Let us
choose $A=[0,\omega]$, $B=[\omega,+\infty]$ to make a concrete
calculation in a specific case, noting that in principle this
corresponds to the computation in Eq.~\eqref{eq:14}.  As represented
in the right panel of Fig.~\ref{fig:3}, this choice of sets leads to
the major simplification
\begin{equation}
	\chi_{B}(\omega_1) = \Theta(\omega_1-\omega)\,,
\end{equation}
where $\Theta(\cdot)$ denotes the Heaviside step function. Moreover, as also shown in the left panel of Fig.~\ref{fig:3}, the resonant manifold is such that $J^{({\rm I})}(\omega;\omega_1,\omega_1-\omega')=0$ for $\omega_1>\omega'$, $J^{({\rm II})}(\omega_p;\omega_1,|\omega'-\omega_1|)=0$ for $\omega_1<\omega'$. Thus, from~\eqref{eq:19} we obtain
\begin{equation}\label{eq:21}
    \mP_{[0,\omega]\rightarrow [\omega,+\infty)} = -2^{d}\pi \int_0^\omega d\omega' \;\omega' \int_\omega^{+\infty} d\omega_1   J^{({\rm II})}(\omega';\omega_1,\omega_1-\omega')\,.
\end{equation}
We are going to derive Eq.~\eqref{eq:14} analytically from
Eq.~\eqref{eq:21}, showing that the {\it Main Statement}~\eqref{eq:5} in Sec.~\ref{sec:2} encompasses the
standard theory of energy fluxes as a particular case.  This proof
relies on the detailed conservation property~\eqref{eq:DC}, from which
it descends that
\begin{equation}\label{eq:23}
\omega_a J(\omega_a;\omega_b,\omega_c) + \omega_b J(\omega_b;\omega_a,\omega_c) + \omega_c J(\omega_c;\omega_a,\omega_b) =0\,,
\end{equation}
for any triad of wavenumbers $\bp, \bp_1, \bp_2$ on the resonant
manifold. An independent proof by construction for the isotropic case
is given in {\it Appendix}~\ref{app:B}. We suggest to the reader to
examine this proof for an intuitive graphical interpretation of
detailed conservation that relies on the symmetries of the resonant
manifold.

\subsection{Proof of the standard flux formula~\eqref{eq:14}}
\begin{quote}{\bf Property: Vanishing self interactions. }
\it The following property holds:
\end{quote}
\begin{equation}\label{eq:24}
	\int_0^\omega d\omega'\int_0^\omega d\omega_1 \; J(\omega',\omega_1, |\omega'-\omega_1|)=0\,,
\end{equation}
This follows directly from detailed conservation, as detailed in {\it Appendix}~\ref{app:A}.

The meaning of this property is that the
integral~\eqref{eq:24} quantifies the flux from $[0,\omega]$ to
$[0,\omega]$, i.e. self interactions that amount to no net transfer of
energy. This leads to the following important corollary of the {\it
  Main Statement}~\eqref{eq:5} 

\medskip
\begin{quote}
\noindent{\bf Retrieving the standard flux formula for isotropic
  systems. } \it The standard flux formula~\eqref{eq:14} that is used
to calculate the energy flux in isotropic wave turbulence is a direct
consequence of Eq.~\eqref{eq:5} ({\it Main Statement}) and
Eq.~\eqref{eq:24}.
\end{quote}
{\bf Proof. } Eq.~\eqref{eq:21} is derived directly from Eq.~\eqref{eq:5}, in the particular case of isotropic systems and adjacent control intervals $A=[0,\omega], B=[\omega,+\infty]$. Exercising the freedom to add zero (i.e. Eq.~\eqref{eq:24}) to Eq.~\eqref{eq:21}, we obtain
\begin{equation}\label{eq:26}
\begin{aligned}
	\mP_{[0,\omega]\rightarrow [\omega,+\infty)} &= -2^{d-1}\pi \int_0^\omega d\omega' \omega'\left[\int_\omega^{+\infty} d\omega_1 J(\omega',\omega_1, |\omega'-\omega_1|) \right.\\
	& \qquad\qquad\qquad\qquad \left.
+ \int_0^\omega d\omega_1 J(\omega',\omega_1, |\omega'-\omega_1|)  \right]\\
	& = -2^{d-1}\pi  \int_0^\omega d\omega' \omega'\int_0^{+\infty} d\omega_1 J(\omega',\omega_1, |\omega'-\omega_1|) \\
	& = \int_0^\omega d\omega' \; \omega' I_{\omega'} = F(\omega)\,,
\end{aligned}
\end{equation}
which concludes the proof of validity of the usual flux
formula~\eqref{eq:14} starting from the {\it Main Statement}~\eqref{eq:5}.$\qquad\square$

\medskip
As highlighted in~\eqref{eq:26}, note that the classical flux expression $F(\omega)$ \eqref{eq:14} contains a self interaction
contribution in the interval $[0,\omega]$. This contribution is vanishing
due to Eq.~\eqref{eq:24}. Moreover,~\eqref{eq:14} requires
regularization at the KZ solution (see
\eqref{eq:16}). Eq.~\eqref{eq:21} is free of such limitations.  We
elaborate on these points in Sec.~\ref{sec:4} by considering surface
capillary waves.

\subsection{Quantifying locality: The transfer integral}\label{sec:TI}

In order to explore the full potential of the {\it Main Statement}~\eqref{eq:5}, let us
introduce a slight generalization of~\eqref{eq:21}. Performing the
outer integration up to a smaller frequency $\tilde\omega<\omega$
allows us to express the power that from the interval
$[0,\tilde\omega]$ is delivered instantaneously to $[\omega,
  +\infty)$:
\begin{equation}\label{eq:27}
    \mP_{[0,\tilde\omega]\rightarrow [\omega,+\infty)} = -2^{d}\pi \int_0^{\tilde\omega} d\omega' \;\omega' \int_\omega^{+\infty} d\omega_1   J^{({\rm II})}(\omega';\omega_1,\omega_1-\omega')\,.
\end{equation}
This is the wave-turbulence analogue of  Eq.~(6.4)
in~\cite{kraichnan1959structure}. Recalling that the collision kernel
has a homogeneity exponent of $\gamma_0-2x$, with a change of variables
$\Omega=\omega'/\omega$, we obtain
\begin{equation}\label{eq:28}
	 \mP_{[0,\tilde\omega]\rightarrow [\omega,+\infty)} = (V_0A)^2 \,\omega^{y+1} \int_{0}^{\tilde\omega/\omega} d\Omega \;T(\Omega)\,,
\end{equation}
where $y=\gamma_0+2-2x$ and
\begin{quote}
{\bf Definition: Transfer Integral.}
\end{quote}
\begin{equation}\label{eq:29}
    T(\Omega):= -2^{d}\pi  (V_0A)^{-2}\;\Omega^{y} \int_{\Omega^{-1}}^{+\infty} d\xi  \; J^{({\rm II})}\left(1;\xi,\xi-1\right)
\end{equation}
\begin{quote}
\it is the {\it transfer integral} of the problem.
\end{quote}
It is a non-dimensional function that captures the inter-scale
``structure'' of the energy transfers between two disconnected regions
of Fourier space. In particular, it quantifies the direct transfer by
a given frequency (smaller than $\omega$) to all frequencies larger than
$\omega$. The integral of $T(\Omega)$ up to $\tilde\omega/\omega$ gives
the distant-transport power exchanged between the two regions
$[0,\tilde\omega]$ and $[\omega,+\infty)$.  Because of the scale
  invariance of the problem, $T(\Omega)$ is uniquely defined no matter
  the chosen values of $\tilde\omega$ and $\omega$. It only has to be
  computed once and then the boundaries of the two sets enter the
  problem as the upper integration boundary and as the scaling factor in Eq.~\eqref{eq:28}.

Using the power $\mP$ between adjacent sets, by using~\eqref{eq:e3.10}-\eqref{eq:28}, we are able to express the Kolmogorov constant of the
problem~\citep{VLvov1992} as a function of the transfer integral itself, for the KZ solution:
\begin{equation}\label{eq:inversion}
	n_p^{KZ} = \kappa_K \sqrt{\mP} \omega^{-x_{KZ}}\,\qquad \kappa_K = \left(V_0 \int_0^1T(\Omega)d\Omega \right)^{-1}\,.
\end{equation}
This inter-scale decomposition of the Kolmogorov constant is one of the important implications of the {\it Main Statement}~\eqref{eq:5}.

How fast $\mP_{[0,\tilde\omega]\rightarrow [\omega,+\infty)}$ tends to
  zero as $\tilde\omega/\omega\to0$ describes how ``local'' or
  ``diffuse'' the energy cascade
  is~\citep{kraichnan1959structure}. This scaling is going to be
  dictated by the asymptotics of $T(\Omega)$, and allows us to improve
  the binary notion of locality (i.e. local/nonlocal) toward a
  more quantitative description. How wide should the separation between
  forcing and dissipation regions be in order to have an inertial
  range sufficiently disconnected from direct interaction with the
  boundaries? The transfer integral $T(\Omega)$ provides a key
  perspective to tackle this type of questions, as will be illustrated
  in the next sections.

\section{Isotropic illustration: surface capillary waves}\label{sec:4}
\subsection{Application of the {\it Main Statement}~\eqref{eq:5}: transfer integral and the Kolmogorov constant}
\begin{figure}
\begin{centering}
\includegraphics[width=\linewidth]{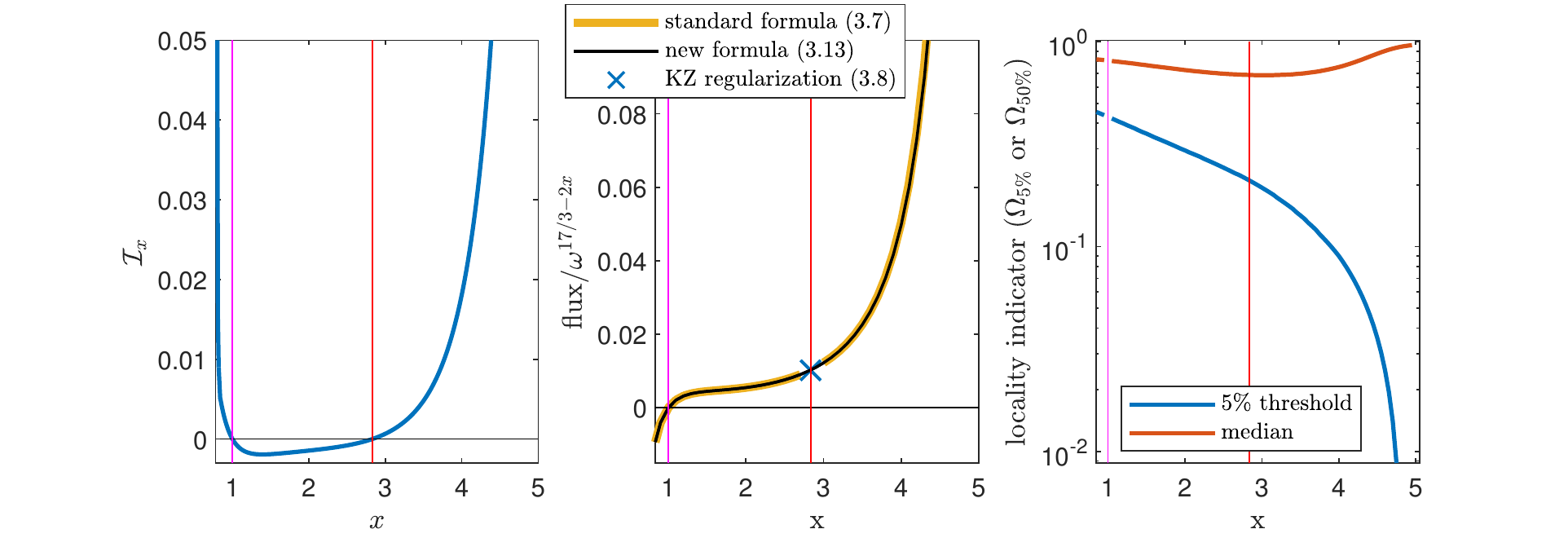}%
  \caption{Left: values of the non-dimensional collision integral
    $\mI(x)$ as a function of the spectral exponent $x$ in the
    locality interval. The two zeros are the RJ and KZ solutions and
    are indicated by a magenta and a red vertical lines,
    respectively. Center: Energy flux normalized by its scaling in
    $\omega$, as a function of $x$. The figure shows perfect agreement
    between the numerical evaluation of the standard flux formula and
    our formula. Moreover, note that the latter does not need to
    be regularized at the KZ solution because it does not contain an
    indeterminate form $0/0$, and the result is identical to the regularization by De~L'Hopital rule. Right: We represent the metrics introduced in formula~\eqref{eq:30} to characterize how local the energy transfers are. All solutions are local for $x\in[5/6,5]$ intended as having an integrable collision operator. However, locality as quantified by~\eqref{eq:30} is stronger for small values of $x$, increasing as $x\to5$ (where $\Omega_{5\%}\to0$).}
 \label{fig:5}
\end{centering}
\end{figure}

Let us consider the problem of surface capillary waves in isotropic
conditions \citep{ZakharovPushkarevPhysicaD}, for which $d=2$. This
system has dispersion relation with $\alpha=\tfrac32$, allowing for
three-wave resonances. After writing the equation in frequency
variables and averaging over the angles in $\bp$-space like
in~\eqref{eq:17}, the interaction kernel has homogeneity exponent
$\gamma_0-2x$, with $\gamma_0=\tfrac83$. Therefore, the stationary states
of the system, in the form $n_\bp=A\omega^{-x}$, are the RJ
equilibrium spectrum with $x=1$ and the KZ spectrum with
$x=(\gamma_0+3)/2=\tfrac{17}{6}$. The convergence conditions of the
collision integral determine the locality interval $[\tfrac56,5]$,
which includes both stationary solutions. For the explicit form of the
WKE we refer the reader to \cite{ZakharovPushkarevPhysicaD}. We
perform the analytical calculations of the locality conditions in {\it
  Appendix}~\ref{app:C}. These calculations are not new per
    se, as they are implied in \cite{ZakharovPushkarevPhysicaD}.  Since we were not able
    to find these calculations in the literature, we included them
    here. In the left panel of Fig.~\ref{fig:5}, we show a numerical
evaluation of the nondimensional collision integral $\mI(x)$,
vanishing in the two stationary states. In the central panel we show
the numerically calculated energy flux between two adjacent sets. This
is done in two ways, according to Eq.~\eqref{eq:14} and to
Eq.~\eqref{eq:21}, showing perfect agreement between the two as proven
analytically in~\eqref{eq:26}. With the precision adopted, the
numerical value so obtained at the KZ solution via
formula~\eqref{eq:21} is identical to the value from the
regularization formula~\eqref{eq:16}, up to a relative error of the
order of $1/1000$. Via the inversion~\eqref{eq:inversion}, this value
relates directly to the Kolmogorov constant of the capillary-wave
problem~\citep{ZakharovPushkarevPhysicaD}. In the most up-to-date
estimates, a direct comparison with the measured flux finds an
agreement within a factor around $1.5-2$ from numerical simulations of
the equations of
motion~\citep{deike2014direct,pan2014direct,pan2017understanding}, and
within a factor of about $3-4$ from
experiments~\citep{deike2014energy}.

Note that the new formula~\eqref{eq:21} can be applied throughout the
locality interval including at the KZ solution, because it does not
contain an indeterminate form~``$0/0$'', as discussed above. Moreover, the
decomposition of the power in terms of the transfer
integral~\eqref{eq:29} is now available also for the KZ solution. We
point out that the integrand of the regularization
formula~\eqref{eq:16}, containing a logarithmic function, is not
equivalent to the transfer-integral
decomposition~\eqref{eq:29}. Indeed, only the latter can be used to
quantify locality and distant-transport scalings.

\subsection{Metrics of locality and distant transport}\label{sec:4.2} %\red{Local versus nonlocal interactions: teleportation
%  versus nearest neighbors?}

We next exploit the formalism of section~\ref{sec:TI} to decompose the energy flux based on the relative separation of the frequencies involved in the transport. We define the quantities  $\Omega_{5\%}$ and $\Omega_{50\%}$ as follows,
\begin{equation}\label{eq:30}
  \int_{0}^{\Omega_{5\%}} d\Omega \;T(\Omega) := 0.05 \int_{0}^{1} d\Omega \;T(\Omega) \,,\quad \int_{0}^{\Omega_{50\%}} d\Omega \;T(\Omega) := 0.5 \int_{0}^{1} d\Omega \;T(\Omega) \,.
\end{equation}
  The first quantity measures the length of the tail of the transfer
  integral.  This quantity therefore indicates how far apart two
  regions in Fourier space have to be for their mutual interactions to be
  negligible. We define ``negligible'' to be five percent of total
  flux of energy.  The second quantity is the median threshold of the
  transfer integral: half of the energy flux is exchanged within this
  threshold range and the other half is exchanged from further than
  this threshold. The right panel of Fig.~\ref{fig:5} shows the
  dependence of $\Omega_{5\%}$ and $\Omega_{50\%}$ on the spectral
  exponent $x$. The median $\Omega_{50\%}$ is always quite close to
  $1$, with a minimum around the KZ solution where
  $\Omega_{50\%}\simeq0.7$. However, the tail metric $\Omega_{5\%}$ is
  decreasing from a value around $0.5$ in the neighborhood of the RJ
  solution, and tends to zero as $x\to5$. In particular, at the
  KZ solution we have $\Omega_{5\%}\simeq0.2$. This means that
  frequencies that are separated by more than half a decade are giving
  a relevant contribution to direct energy transport in the KZ
  stationary state!

\begin{figure}
\begin{centering}
\includegraphics[width=\linewidth]{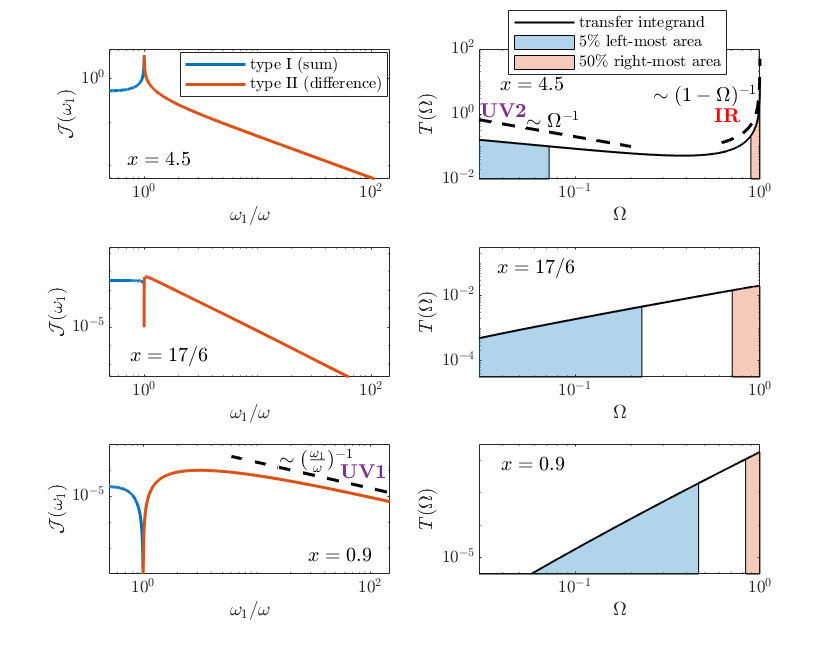}%
  \caption{Interaction kernel and transfer integral for three different values of $x$. All plots are in $\log-\log$ scale. The three solutions here represented are ``local'': the interaction kernel is integrable. The limiting scalings of the three locality conditions IR, UV1 and UV2 (cf Sec.~\ref{sec:7}) are represented by the black dashed lines.}
 \label{fig:6}
\end{centering}
\end{figure}

The detail of the transfer integral calculations are shown in
Fig.~\ref{fig:6}, for three different values of $x$: $4.5$,
$\tfrac{17}{6}$ and $0.9$, from top to bottom. The left panels show
the magnitude of the interacion kernel. In the first two cases, the
type I contributions are positive and the type II negative,
corresponding to a direct cascade. In the latter case, the signs are
exchanged, corresponding to inverse cascade for $x<1$.

The singularity in $1$ for large values of $x$ behaves like
$|\omega_1/\omega-1|^{-x+3}$. With the two integrations in
Eq.~\eqref{eq:27}, the convergence condition must be $x<5$, retrieving
the infrared (IR) locality condition. This implies that the transfer integral is
dominated by an integrable singularity $T(\Omega)\simeq
(1-\Omega)^{-x+4}$ for $\Omega\to1$, when $x>4$.

The scaling of the interaction kernel for $\omega_1/\omega\gg1$ is given by $(\omega_1/\omega)^{-x-1/6}$ for $x\simeq1$, and by $(\omega_1/\omega)^{-x-1/3}$ for $x\gg1$. For the transfer integral $T(\Omega)$ to converge it must be $-x-1/6<-1$, which gives the familiar ultraviolet (UV)
locality condition $x>5/6$. Notice the proximity of the case $x=0.9$ to this limit scaling in the bottom left panel of Fig.~\ref{fig:6}. By Eq.~\eqref{eq:29}, this implies an asymptotic scaling $T(\Omega)\simeq {\Omega^{4-x}}$ for $\Omega\ll1$. This will be discussed further in Sec.~\ref{sec:7}.%,
                                                                                                                                                                                                                                                                                                                                                                                                                                                                                                                           %when
                                                                                                                                                                                                                                                                                                                                                                                                                                                                                                                           %$x<3$.

Let us use this result to estimate the asymptotic scaling of the
distant-transport power:
\begin{equation}\label{eq:4.2}
	\mP_{[0,\tilde\omega]\rightarrow [\omega,+\infty)} = \int_0^{\tfrac{\tilde\omega}{\omega}} T(\Omega)d\Omega \sim \left(\frac{\tilde\omega}{\omega}\right)^{5-x}\,,\quad \text{as }\; \frac{\tilde\omega}{\omega} \to 0\,.
\end{equation}
For the KZ solution, $x=17/6$, this yields $\mP_{[0,\tilde\omega]\rightarrow [\omega,+\infty)} \sim \left(\tfrac{\tilde\omega}{\omega}\right)^{13/6}$.

Thus, the energy cascade at the KZ stationary solution of capillary
waves can be considered quite strongly local; moreover, the energy
transport for spectra that are steeper than KZ becomes more and more
diffuse, while for whiter spectra it becomes more and more local
(cf. Fig.~\ref{fig:5}). Around equilibrium, $x\simeq1$, the scaling decay is
$\left(\tfrac{\tilde\omega}{\omega}\right)^{23/6}$. The analysis
presented here suggests a viable approach to quantifying how far from
the dissipation and forcing regions one should be in order for direct
energy transfers with the boundaries to be fairly negligible. Taking
the $\Omega_{5\%}$ as a reasonable (albeit arbitrary) cutoff for a
notion of ``negligibility'', for KZ we would obtain at least a factor
of $5$ of separation from each boundary. A criterion of this sort
would exclude about $1.4$ orders of magnitude (half on each side) from
being a part of an inertial range fairly independent of both the
forcing and the dissipation regions. For surface capillary waves,
which are constrained on scales from $0.5$ mm to $17$ mm, there are
about $2.3$ orders of magnitude of available frequencies, which is not much larger than $1.4$. We refer the reader to Sec.~\ref{sec:7} for further discussion.

These and similar quantifications of fluxes and associated level of
locality are directly applicable to any wave-turbulence system. They
open the possibility to an analysis of energy transfers that goes
beyond the mere stationary states to explore transients, boundary
effects, and a scale-by-scale decomposition of the energy transfer
contributions.

%The transfer integral $T(\Omega)$ is computed on the type II branch, Eq.~\eqref{eq:29}.

\section{Anisotropic systems}\label{sec:5}
\subsection{Overview on the theory of energy fluxes}\label{sec:3.2.1}
A direct extension to anisotropic systems of the theory of energy
fluxes reviewed in Sec.~\ref{sec:3.1.1} is possible, in
scale-invariant and stationary conditions. It consists of the use of
generalized Zakharov-Kraichnan-Kuznetsov conformal
transformations~\citep{kuznetsov1972} to find generalized KZ solutions
\citep{VLvov1992,NazBook}. Each of these solutions corresponds to the
stationary cascade solution of one of the positive-definite conserved
quantities of the WKE. In principle, each of these positive invariants
also corresponds to an independent equilibrium solution. However, at
variance from the isotropic case, both these types of equilibrium and
nonequilibrium stationary solutions are not the only possible
stationary solutions, but only particular ones. In the case of 3D
systems with two effective independent dimensions, there are two
families of an infinite number of equilibrium and nonequilibrium
solutions, respectively represented by the points of two 1D curves in
the 2D plane of possible power-law exponents. Physical examples are
the Rossby/drift waves, where there are three positive collision
invariants and three KZ solutions~\citep{balk1990nonlocal,NazBook}, or
the internal gravity waves, where there is one known collision
invariant (the energy) and one corresponding KZ solution~\citep{pelinovsky1977weak,LT}.
One subsequent necessary step in the theory is the verification that
these stationary solutions correspond to a convergent collision
integral, i.e. that they are local.  For internal gravity waves, for
instance, there is only one local stationary solution which is found
numerically~\citep{iwthLPTN} and is different from the KZ solution.

The calculation of the fluxes for the anisotropic KZ solutions leads
to a regularization similar to Eqs.~\eqref{eq:14}-\eqref{eq:16}. Here,
we make use of simple-minded dimensional analysis to illustrate its
properties. For concreteness, we will use the example of horizontally
isotropic internal gravity waves. In
the hydrostatic approximation, the scale-invariant dispersion relation
and matrix elements read~\citep{olbers1976nonlinear,LT2}
\begin{equation}\label{eq:32}
	\omega_{\bp} = \gamma \frac{k}{m}\,,\qquad
        \left|V_{12}^0\right|^2=V_0^2 k^{2\mu_k}m^{2\mu_m}
        f\left(\frac{k_1}{k},\frac{k_2}{k},\frac{m_1}{m},\frac{m_2}{m}\right)\,,
\end{equation}
where $\gamma$ and $V_0$ are dimensional constants, and $k$ and $m$ are the magnitude of the (2D-) horizontal and (1D-) vertical wavenumbers, respectively. Moreover, we have $\mu_k=3/2$ and $\mu_m=-1/2$. In an inertial range where no external forcing or dissipation are present, we look for general stationary solutions for the action density of the form
\begin{equation}\label{eq:33}
	n_\bp=A k^{-a} m^{-b}\,.
\end{equation}
Let us study energy propagation in the positive quadrand
$k-m$, by defining the horizontally-averaged wave action $n(k,m)= 4\pi
k n_\bp$ and energy $e(k,m)=\omega_\bp n(k,m)$. The standard use of
the differential conservation equation for energy yields
\begin{equation}\label{eq:34}
	\frac{\partial e(k,m)}{\partial t} = 4\pi k \omega_\bp I_\bp = -\frac{\partial F_k(k,m)}{\partial k} - \frac{\partial F_m(k,m)}{\partial m}\,,
\end{equation}
where the dependence on time is implicit, $I_{\bp}$ is the collision integral, and $F_k$ and $F_m$ are the horizontal and vertical components of the energy flux in $k-m$ space. Using Eqs.~\eqref{eq:32}-\eqref{eq:33}, from dimensional analysis we obtain
\begin{equation}\label{eq:35}
	k \omega_\bp I_\bp = (V_0 A)^2 k^{6-2a} m^{-2b}\mI(a,b)\,,
\end{equation}
where $\mI(a,b)$ is the non-dimensional collision integral that
vanishes in the stationary states.  Let us plug this into the
r.h.s. of~\eqref{eq:34}. We follow a similar reasoning as for
Eq.~\eqref{eq:14} in the isotropic case.   Let us assume that $F_k=F_k(m)$ and $F_m=F_m(k)$ -- a-posteriori, the KZ solution is found to enjoy this property. Under this assumption, we integrate in $k$ from $0$ to $k$ to
obtain the horizontal component of the flux and in $m$ from $0$ to $m$ to obtain the
vertical component. We obtain:
\begin{equation}\label{eq:36}
	F_k = -4\pi(V_0 A)^2 \frac{k^{7-2a}m^{-2b}}{7-2a}\mI(a,b)\,,\;\; F_m =-4\pi (V_0 A)^2\frac{k^{6-2a}m^{1-2b}}{1-2b}\mI(a,b)\,.
\end{equation}
Using the generalized Zakharov-Kraichnan-Kuznetsov transformations,
one finds out that the generalized KZ solution is the particular one
for which the above denominators vanish~\citep{LT}, setting $a_{\rm
  KZ}=7/2,b_{\rm KZ}=1/2$, the Pelinovski-Raevski
spectrum~\citep{pelinovsky1977weak}. Because for this solution also
the numerators vanish, in analogy with~\eqref{eq:16}, one can
regularize the $0/0$ indeterminate form by de L'H\^opital's rule to obtain~\citep{VLvov1992}:
\begin{equation}\label{eq:37}
	F_k^{\rm KZ} = 2\pi (V_0A)^2{m^{-1}}\left.\frac{\partial\mI}{\partial a}\right|_{(a_0,b_0)}\,,\qquad F_m^{\rm KZ} =2\pi (V_0A)^2{k^{-1}}\left.\frac{\partial \mI}{\partial b}\right|_{(a_0,b_0)}\,.
\end{equation}

Notice that the KZ solution is the particular case for which
        the $k$-component is independent of $k$, i.e. $F_k=F_k(m)$,
        and the $m$-component is independent of $m$,
        i.e. $F_m=F_m(k)$. This spectrum
    is known to be
nonlocal~\citep{iwthLPTN}. In particular, the point $(\tfrac72,\tfrac12)$ in the
$a-b$ plane exhibits divergencies at both high
    and low wavenumbers. These divergencies can also be seen from
Eq.~\eqref{eq:37}: integrating the flux along any boundary in $k-m$
space yields a logarithmic divergence both at low wavenumbers
($k\to0,m\to0$) and at large
wavenumbers ($k\to\infty,m\to\infty$). Thus, the equations~\eqref{eq:37} can be written only in a formal way, but in practice they have no meaning. It was shown in~\cite{iwthLPTN} that
        there is only one stationary solution that is local, with
        spectral exponent values $a=3.69,b=0$. However, for this non-KZ stationary solution, Eq.~\eqref{eq:34} is merely stating that the
        divergence of the flux is zero.  Therefore, it is only possible to determine the direction of the flux, but the magnitude of the flux of energy remains undetermined. This is shown in {\it  Appendix~\ref{DoesNotWork}}.

\section{Anisotropic illustration: Internal gravity waves}\label{sec:6}

\subsection{Definition of the problem}

Here, we illustrate an application of formula~\eqref{eq:5} to the
anisotropic problem of internal gravity
waves~\citep{olbers1976nonlinear,caillol2000kinetic,LT2}. This is
nontrivial in several ways, involving: (i) physically motivated
control volumes in Fourier space with a geometry that is more interesting than simple rectangles in $k-m$
space; (ii) a relevant stationary spectrum that is not a KZ solution (the solution $a=3.69,b=0$);
(iii) a renowned spectrum (the Garrett and Munk spectrum) that is not
stationary under the wave kinetic equation evolution; (iv) decomposition of the
fluxes in terms of transfer integrals allowing to quantify the level
of locality. Each of these these
points could not be analyzed fully by the standard wave turbulence
theory of energy fluxes summarized in Section~\ref{sec:5}.
    These results were obtained
    intuitively in
    \cite{dematteis_lvov_2021} and \cite{dematteis2022origins}. Here,
    we provide firm mathematical justification for results of such
    type, study locality of internal wave interactions and analyze the
    celebrated
    Garrett and Munk spectrum.

\begin{figure}
\begin{centering}
\includegraphics[width=0.7\linewidth]{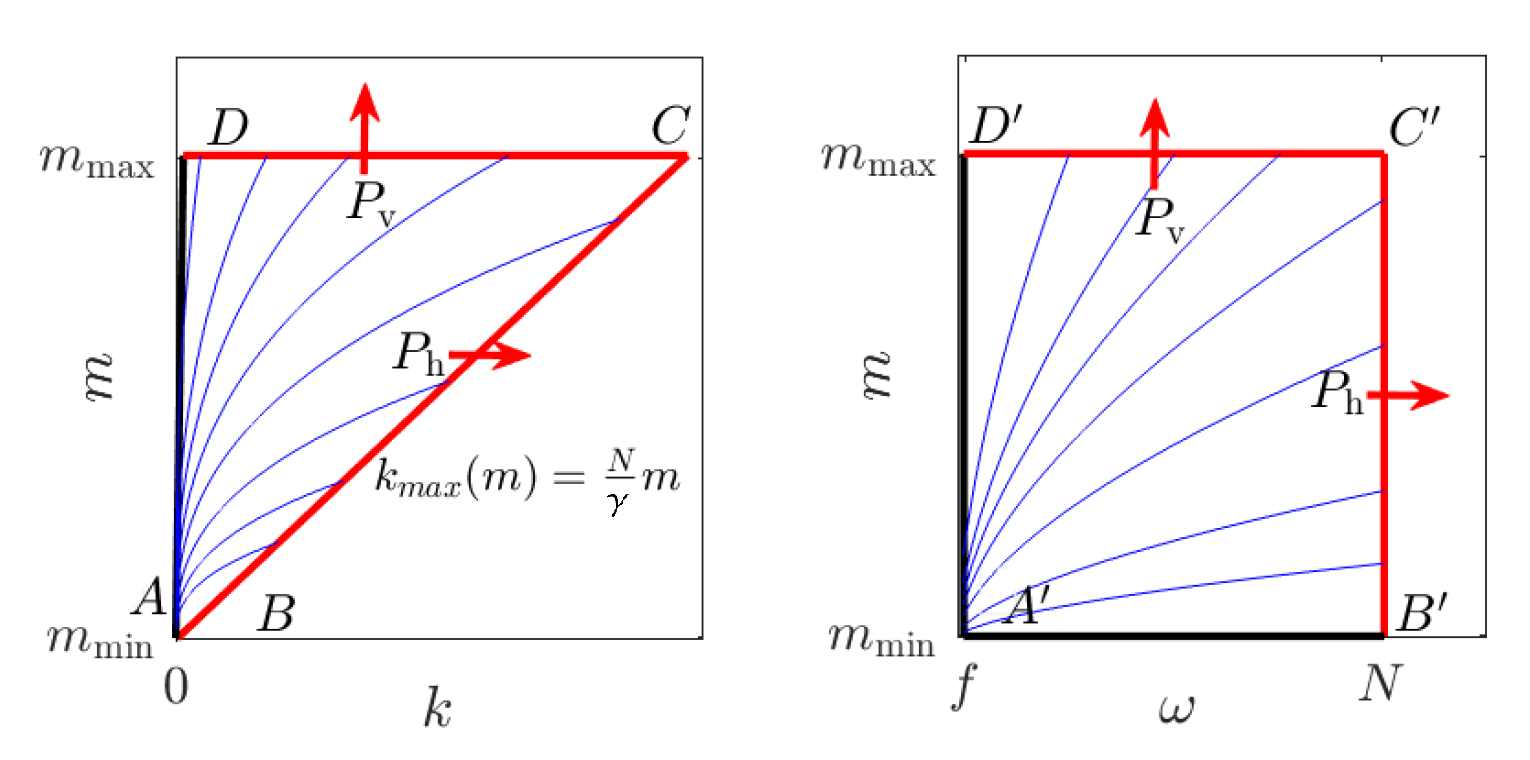}%
\caption{Representation of the 2D Fourier space for the anisotropic
  internal-wave problem. Left: in horizontal-vertical wave
      number $k-m$ coordinates; Right: in frequency-vertical
      wavenumber $\omega-m$ coordinates. The streemlines represent
  the energy flux vector field and are drawn from Eq.~\eqref{eq:38}
  for the stationary solution $a=3.69,b=0$. Eq.~\eqref{eq:38} is
  determined up to an arbitrary factor $C$, but this nonetheless
  allows us to know the flux direction (thus, the streamlines). The
  change of coordinates from the $k-m$ to the $\omega-m$
  representation is given by the dispersion
  relation~\eqref{eq:32}. The quantities $P_{\rm h}$ and $P_{\rm v}$
  are computed rigorously in Eq.~\eqref{eq:39}. The computations in
  this paper are performed in $k-m$ space, although the physical
  boundaries are naturally defined in $\omega-m$ space. This figure
  illustrates the equivalence between the two representations.}
 \label{fig:7}
\end{centering}
\end{figure}

The boundaries in
    spectral space are naturally defined for internal waves.
The frequency $\omega$ takes values in the interval $[f,N]$, where $f$
and $N$ are the inertial and the buoyancy frequencies, respectively.
These two frequencies give the minimal and maximal frequencies
($f\ll N$) of the problem with the {\it inertial range} between them.
The vertical wavenumber $m$ takes values in
$[m_{\rm min},m_{\rm max}]$, where
$m_{\rm min}=2\pi/H, m_{\rm max}=2\pi/h$, with $H$ and $h$ being the
ocean depth and the vertical scale past which internal waves become
unstable due to shear instability. Let us
assume that the box $\mA=[f,N]\times[m_{\rm min},m_{\rm max}]$ is the
inertial range, and for $m>m_{\rm min}$ and $\omega>N$ strong
turbulence acts as an idealized sink. Suitable energy sources will
indeed be necessary at the bottom and left boundaries of the
``inertial box'' $\mA$, in order for an energy cascade to be sustained
in time. The inertial box $\mA$ is shown in Fig.~\ref{fig:7}, both in
$k-m$ space and in $\omega-m$ space. The change of variables between
the two spaces is prescribed by the dispersion
relation~\eqref{eq:32}. In Fig.~\ref{fig:7}, we also show the
streamlines of the energy flux obtained by dimensional arguments in {Appendix~\ref{DoesNotWork}}, Eq.~\eqref{eq:38},
for the stationary state with $a=3.69,b=0$.
These lines give a sense of the need for a source at low frequencies
and low wavenumbers for energy to be delivered to the whole inertial
box. We use $\mA$ as our input control volume. For the output control
volume, $\mB$, we consider two possibilities: either
$\mB_{\rm h}=\{(\omega,m): \omega>N\}$, or
$\mB_{\rm v}=\{(\omega,m): m>m_{\rm max}\}$. In the first case, the
power $\mP_{\mA\rightarrow\mB}$ defines the quantity $\mP_{\rm h}$,
the instantaneous power transferred ``horizontally'' through the
boundary denoted as $BC$ in Fig.~\ref{fig:7}. In the second case,
$\mP_{\mA\rightarrow\mB}$ defines the quantity $\mP_{\rm v}$, the
instantaneous power transferred ``vertically'' through the boundary
denoted as $CD$. The powers $P_{\rm h}$ and $P_{\rm v}$ are calculated rigorously in
the next section using the {\it Main Statement}~\eqref{eq:5}.

\subsection{Application of the {\it Main Statement}~\eqref{eq:5} and transfer integrals}\label{sec:6.2}
Applying
formula~\eqref{eq:5}, we obtain \citep{dematteis_lvov_2021}:
\begin{equation}\label{eq:39}
	 \mP_{\rm h} =  \int_{m_{\rm min}}^{m_{\rm max}} d m \;\mF_{\rm h}(m)\,, \quad \mP_{\rm v} = \int_{\frac{f}{\gamma}m_{\rm max}}^{\frac{N}{\gamma}m_{\rm max}} d k \;\mF_{\rm v}(m)\,,
\end{equation}
where
\begin{equation}\label{eq:40}
	 \mF_{\rm h}(m) =\frac{N^2}{g}(V_0A)^2 \left(\frac{N}{\gamma}m\right)^{7-2a}m^{-2b} \int_{\frac{f}{N}}^{1} d K \;T_{\rm h}(K)\,,
\end{equation}
\begin{equation}\label{eq:41}
    T_{\rm h}(K)= -\frac{4\pi }{ (V_0A)^{2}}\;K^{6-2a} \int d\xi_1d\xi_2  \; \sum_l \chi_{\rm K}^{(l)}(\xi_1,\xi_2)\;J^{({ l})}\left(\xi=1, \mu=1;\xi_1,\xi_2\right)
\end{equation}
and
\begin{equation}\label{eq:42}
	 \mF_{\rm v}(k) =\frac{N^2}{g}(V_0A)^2 k^{6-2a} m_{\rm max}^{1-2b} \int_{\frac{m_{\rm min}}{m_{\rm max}}}^{1} d M \;T_{\rm v}(M)\,,
\end{equation}
\begin{equation}\label{eq:43}
    T_{\rm v}(M)= -\frac{4\pi  }{(V_0A)^{2}}\;M^{-2b} \int d\xi_1d\xi_2  \; \sum_l \chi_{M}^{(l)}(\xi_1,\xi_2)\;J^{({ l})}\left(\xi=1, \mu=1;\xi_2,\xi_2\right).
\end{equation}
Here, $J^{({ l})}\left(\xi=1, \mu=1;\xi_1,\xi_2\right)$ denotes the
six resonant branches of the interaction kernel corresponding to the
triad of non-dimensional horizontal wavenumbers
$\xi=1,\xi_1=k_1/k,\xi_2=k_2/k$, and vertical wavenumber $\mu=1$. The six
resonant conditions determine the values of $\mu_1=m_1/m$ and
$\mu_2=m_2/m$. The characteristic interaction weights $\chi_K^{({
    l})}(\xi_1,\xi_2)$ are defined by the
rules in Table~\ref{tab:1}, taking $\mB_{\rm
  h}=\{\xi:\xi>K^{-1}\}$, with $K^{-1}>1$. The weights $\chi_M^{({
    l})}(\xi_1,\xi_2)$ are defined likewise by taking $\mB_{\rm
  v}=\{\mu:\mu>M^{-1}\}$, with $M^{-1}>1$, where the condition $\mu>M^{-1}$ is
applied to the non-dimensional vertical wavenumbers $\mu_1$ and
$\mu_2$ found as solution of the $l-$th resonant branch.

Using similar equations, it was shown in \cite{dematteis_lvov_2021}
and \cite{dematteis2022origins} that both the scaling and the
prefactor of the total calculated power are in agreement with the
observational finescale parameterization of oceanic turbulent
production~\citep{polzin2014finescale}, up to a factor $1.5$
difference. In a loose sense, these calculations are equivalent to
    evaluating the Kolmogorov constant for the internal wave
problem, i.e. expressing the explicit theoretical relationship between the energy flux and the spectral energy density.

\subsection{Metrics of locality and distant transport}

\begin{figure}
\begin{centering}
\includegraphics[width=\linewidth]{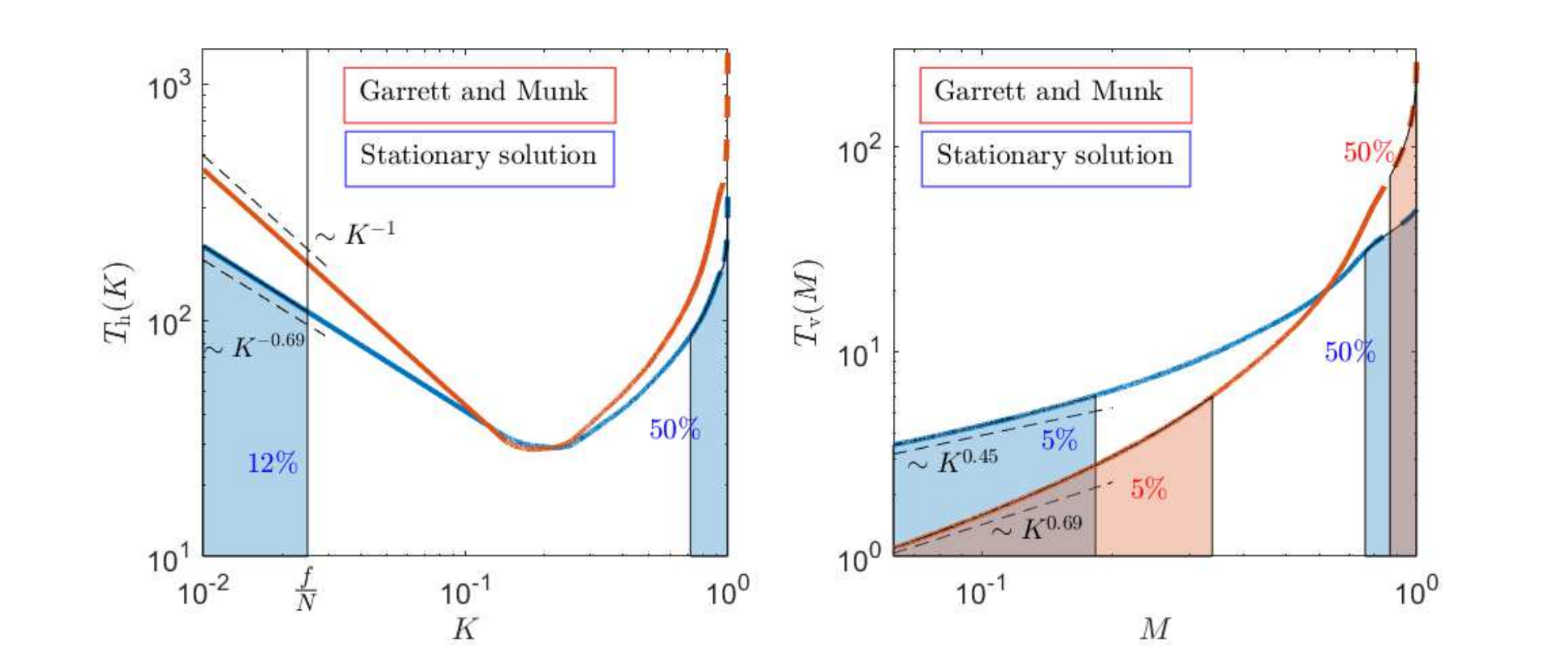}%
\caption{Transfer integrals of internal gravity waves for horizontal
  transport (left, Eq.~\eqref{eq:41}) and vertical transport (right,
  Eq.~\eqref{eq:43}). For the power-law spectrum~\eqref{eq:33}, the
  stationary solution has exponents $a=3.69,b=0$, and the scale
      invariant limit of the Garrett and Munk spectrum has exponents
  $a=4,b=0$.}
 \label{fig:8}
\end{centering}
\end{figure}

The methodology developed in this manuscript allows not only to compute the fluxes
    of energy, but also to analyze the locality of interactions. In
relation to their isotropic version~\eqref{eq:28}-\eqref{eq:29},
indeed the expressions~\eqref{eq:39}-\eqref{eq:43} feature one extra
integration along the boundaries of the 2D inertial box. The
quantities $\mF_{\rm v}(k)$ and $\mF_{\rm h}(m)$ are energy fluxes per
unit of $k$ and $m$, respectively. Thanks to scale invariance, their
dependence in $k$ and $m$ is given by the scaling relations
in~\eqref{eq:40} and~\eqref{eq:42}. Therefore, to study level
    of locality of the interactions it is sufficient to study
$\mF_{\rm v}(k=1)$ and $\mF_{\rm h}(m=1)$, whose structure is
expressed in terms of the transfer integrals $T_{\rm h}$ and
$T_{\rm v}$ in~\eqref{eq:41} and~\eqref{eq:43}. This is represented in
Fig.~\ref{fig:8}. Both for $T_{\rm h}$ and $T_{\rm v}$, the further
from the boundary at $K=1$ or $M=1$, the more nonlocal (i.e., with
large scale separation) the contribution to the energy transfer. The
dashed colored lines on the right of each panel indicate the
analytical leading orders (integrable singularities) from the IR
region of the resonant manifold. The scalings on the left side are
given by the UV leading orders, multiplied by the factor
in~\eqref{eq:41} and~\eqref{eq:43}. The shaded areas indicate the
left-most and the right-most contributions to the total flux, in a
percentage amount indicated in the figure. Two different spectra in
the form~\eqref{eq:33} are studied 
    below~\citep{iwthLPTN}.
\begin{itemize}
\item $a=3.69, b=0$ (stationary state of the internal WKE)

  For the horizontal transport, despite having the right-most $50\%$
  coming from the $[0.7,1]$ interval (i.e. the median is about $0.7$),
  the heavy tail implies that about $12\%$ of the power $\mP_{\rm h}$
  is transferred directly from modes that are smaller than the left
  boundary of the inertial box, i.e. $\omega<f$ -- if we take
  $N/f=40$, a realistic oceanic aspect ratio. Since there are no waves at
  $\omega<f$, this is not possible. Nevertheless, it indicates that
  the horizontal transfer is highly nonlocal -- even though the
  spectrum is ``local'' in terms of convergence of the collision
  integral. Even the lowest frequencies in the system are
  energetically connected with the dissipation region at high frequency in a
  non-negligible way.

For the vertical transport, the situation is much more local. The
median is around $0.8$, and the right-most $5\%$ of the energy tranfer comes from the
left of approximately $0.2$. Because a realistic range of vertical
scales varies by a factor
of the order of $200$, a factor of $5$ of distance from the
boundary at $m_{\rm max}$ is relatively quite small. Thus, the vertical
transport is highly local.

\item $a=4, b=0$ (scale-invariant limit of the Garrett and
  Munk spectrum)

The horizontal transport power is marginally divergent, due to a
$K^{-1}$ singularity as $K\to0$. Therefore, it is not meaningful to
indicate percentage metrics of the contribution.

The vertical transport is highly local: more so than for the
stationary spectrum. About $95\%$ of the total power $\mP_{\rm v}$
comes from the region within a factor of $3$ from the dissipation boundary.
\end{itemize}

Let us exploit the transfer integrals to define the distant-transport
fluxes, neglecting the dimensional prefactors
\begin{equation}\label{eq:44}
\begin{aligned}
	& \mF_{{\rm h},[0,\tilde\omega]\rightarrow [\omega,+\infty)}
    \propto \int_{0}^{\tilde\omega/\omega} dK \;T_{\rm h}(K)\,,\\ &
    \mF_{{\rm v},[0,\tilde m]\rightarrow [m,+\infty)} \propto
      \int_{0}^{\tilde m/m} d M \;T_{\rm v}(M)\,.
\end{aligned}
\end{equation}
For horizontal transport, we know the analytical scaling
$T_{\rm h}(K)\propto K^{3-a}$ as $K\to0$ (cf. left panel of
Fig.~\ref{fig:8}). This means that we have:
$\mF_{{\rm h},[0,\tilde\omega]\rightarrow [\omega,+\infty)}\propto
\left(\tfrac{\tilde\omega}{\omega}\right)^{4-a}$, as
$\tfrac{\tilde\omega}{\omega}\to0$. This shows that the flux for the
Garrett and Munk spectrum is marginally divergent
(logarithmic divergence) and that the stationary spectrum has a very
weak decay with scaling
$\left(\tfrac{\tilde\omega}{\omega}\right)^{0.31}$. For vertical
transport, we use the numerical scalings shown in the right panel of
Fig.~\ref{fig:8}. These imply for
$\mF_{{\rm v},[0,\tilde m]\rightarrow [m,+\infty)}$ a scaling
$ \left(\tfrac{\tilde m}{m}\right)^{1.69}$ for the Garrett and Munk
spectrum, and a scaling
$ \left(\tfrac{\tilde m}{m}\right)^{1.45}$ for the stationary
spectrum. %In both cases this is a fairly strong decay, comparable
    %with the one found by~\cite{kraichnan1959structure} for 3D
    %turbulence with exponent $3/2$.

We end the section by showing that even when a solution is
mathematically {\it nonlocal}, a regularization takes place if one
considers physical cutoffs. For the high wavenumber limit of
    the Garrett and Munk spectrum, indeed a fairly plausible
oceanic condition~\citep{le2021variability,pollmann2020global,thakur2022impact}, the
integral defining the horizontal energy flux has a logarithmic
divergence (considering an idealized zero minimal frequency).
However, real physical systems have boundaries and other constraints
that imply natural lower and upper cutoffs in Fourier space. For
instance, oceanic internal waves cannot oscillate at frequencies lower
than the Coriolis frequency $f$. Imposing this lower cutoff by hand,
the horizontal flux in Eq.~\eqref{eq:44} is given by
\begin{equation}\label{eq:50}
	\mF_{{\rm h},[f,\tilde\omega]\rightarrow [\omega,+\infty)} \propto \int_{f/\omega}^{\tilde\omega/\omega} dK \;K^{3-a}\propto \frac{1}{4-a}\frac{\tilde\omega^{4-a} - f^{4-a}}{\omega^{4-a}}\,,
\end{equation}
For a given frequency $\tilde\omega>f$, the flux is finite and
continuous in $a$, with
$\lim_{a\to4^{\pm}}\mF_{{\rm h},[f,\tilde\omega]\rightarrow
  [\omega,+\infty)}=|\log(f/\omega)|$.  For $a\ll4$, the contribution
to the flux from $[f,\tilde\omega]$ is concentrated around
$\tilde\omega$. As $a\to4^-$, the contribution becomes less and less
concentrated in $\tilde\omega$. For $a>4$, the contribution becomes
more concentrated in $f$ than in $\tilde\omega$, and increasingly so
as $a$ increases. Since the nonlocality and boundary dependence are
smooth functions of $a$, and the transition is continuous in $a=4$, it
is clear that interpreting this threshold as a sharp definition of
physical realizability/non-realizability is quite fictitious. For a
physical system with a finite available range of scales, highly
nonlocal spectra will indeed show strong dependence on the boundaries,
as Eq.~\eqref{eq:50} demonstrates.
If the forcing is varying strongly in time, for instance,
this will correspond to transient forcing-driven conditions, with the
system being highly influenced by the forcing variability at the lower
boundary of Fourier space. However, it may still be of fundamental
importance to quantify the associated energy fluxes: the energy fluxes
associated to highly nonlocal and transient spectra of internal waves
play a crucial role for the oceanic circulation and climate at
large~\citep{polzin2014finescale}.

\section{Integrability conditions and locality of energy transport}\label{sec:7}
        
            Here, we consider the conditions for a finite energy transfer and discuss their consistency with the standard locality conditions
of wave turbulence~\citep{VLvov1992}. Moreover, we suggest a way to quantify the locality properties of a wave-turbulence spectrum.

Consider a generic isotropic system with
distant-transport power described
by~\eqref{eq:28}-\eqref{eq:29}. Defining $f(\xi)=J^{({\rm
    II})}\left(1;\xi,\xi-1\right)$, assume the following definitions
for the scaling exponents $\gamma_0,\gamma_1,\gamma_2, y$,
\begin{equation}\label{eq:45}
\begin{aligned}
	&J^{({\rm II})}\left(k;k_1,k_1-k\right) = k^{\gamma_0-2x}f\left(\frac{k_1}{k}\right)\,,\qquad y = \gamma_0+2-2x\,,\\
	&f(\xi) \propto \xi^{\gamma_2-x} \text{ as }\xi\to+\infty\,,\qquad  f(\xi)\propto (\xi-1)^{\gamma_1-x} \text{ as }\xi\to1^+\,. \\
\end{aligned}
\end{equation}
For the distant-power~\eqref{eq:28} to be finite for any
$\tilde\omega\in[0,\omega]$, we have to impose the integrability of
$T(\Omega)$ defined by~\eqref{eq:29}, in the integration domain
$[0,\omega]$. Setting the prefactor to unity for simplicity, we recall
that
\begin{equation}\label{eq:46}
 \mP_{[0,\omega]\rightarrow [\omega,+\infty)} = \omega^{y+1} \int_{0}^{1} d\Omega \;T(\Omega)\,, \qquad  T(\Omega)= \Omega^{y} \int_{\Omega^{-1}}^{+\infty} d\xi  \; f(\xi)\,.
\end{equation}
Integrability for $\Omega\to1$, due to double integration, gives the
\cite{VLvov1992} IR condition
\begin{equation}\label{eq:47}
 x<\gamma_1+2\,.
\end{equation}
Integrability of $f(\xi)$ for $\xi\to+\infty$ gives the \cite{VLvov1992} UV condition (UV1)
\begin{equation}\label{eq:48}
 x>\gamma_2+1\,.
\end{equation}
Now, we have a third condition from integrability of $T(\Omega)$ for $\Omega\to0$. This is also a UV condition (UV2) and reads
\begin{equation}\label{eq:49}
 x<2+\gamma_0-\gamma_2\,.
\end{equation}
For a scale invariant spectrum $n_\bp=A\omega^{-x}$, the instantaneous
power exchanged between the sets $[0,\omega]$ and $[\omega,+\infty)$
is finite if and only if the three conditions~\eqref{eq:47}
(IR),~\eqref{eq:48} (UV1) and~\eqref{eq:49} (UV2) are simultaneously
fulfilled.  This result needs to be compared with the standard {\it
  locality conditions} of wave turbulence
theory~\citep{VLvov1992,NazBook}, which consist of Eqs.~\eqref{eq:47}
(IR) and~\eqref{eq:48} (UV1). Let us use the example of the
capillary-wave system, where we have $\gamma_1=3,\gamma_0=8/3$,
$\gamma_2=-1/6$ for $x<1$ and $\gamma_2=-1/3$ for $x>1$. We provide a
detailed calculation of these scalings in {\it
  Appendix}~\ref{app:C}. The three convergence conditions give
$x<5,x>5/6, x<5$, respectively. The three conditions are represented
by the three black dashed lines in Fig.~\ref{fig:6}. As we see,
    for the case of capillary waves, the
third condition is identical to the first one, showing that the
integrability interval computed by imposing a finite energy transport
is fully consistent with the usual locality
conditions. Therefore, for the wave turbulence
    spectrum of any wave turbulence system to be truly local, all of
    these three locality conditions need to be individually checked and verified.

Using the quantities defined in Sec.~\ref{sec:4.2}, we propose the
definition of a nondimensional number that quantifies the width of
direct energy transport in Fourier space:
\begin{equation}\label{eq:7.7}
	w = -\log_{10}\Omega_{5\%}\,,
\end{equation}
in units of orders of magnitude.
    This quantity is a measure of the inter-scale width of the resonant interactions. If, for example, this quantity tends to zero, it means that the interactions are highly local. If this quantity is comparable to the range of scales that are physically available, there cannot exist an inertial range of scales where the interactions are sufficiently independent of the boundaries. This quantity diverges for spectra outside the locality interval.
Using the power-law tail scaling $T(\Omega)\sim c\Omega^{1+\gamma_0-\gamma_2-x}$ as $\Omega\to0$, and the definition~\eqref{eq:30}, we obtain the estimate
\begin{equation}
	w = (x-2-\gamma_0+\gamma_2) \log_{10}\left(\tfrac{0.05}{c}(2+\gamma_0-\gamma_2-x)\int_0^1 T(\Omega)d\Omega\right)\,.
\end{equation}
In the examples illustrated in this manuscript (cf. Figs.~\ref{fig:6}
and~\ref{fig:8}), we have $w\simeq 0.3$ near equilibrium and
$w\simeq 0.7$ at the KZ solution of surface capillary waves. For the
horizontal transport of internal waves, we have $w\simeq2.1$ for the
stationary solution and $w\to\infty$ for the Garrett and Munk scale
invariant limit (logarithmic divergence). For vertical transport, we
have $w\simeq0.7$ for the stationary solution and $w\simeq0.5$ for the
Garrett and Munk spectrum of internal waves. We can see that $w$ is finite if $x$ is in the locality
    interval. However, its value can vary from close to zero, when
    transport is highly local, to fairly large values even if the
    spectrum is ``local'' (see e.g. horizontal transport for internal
    waves in Sec.~\ref{sec:6}). The estimate of $w$ can be important
    for understanding how wide the inertial range of wave turbulence
    must be in the experiments, in order to become independent of the
    boundaries.

\section{Discussion}\label{sec:8}

We introduced formula~\eqref{eq:5} for systematic computation of any
inter-scale energy transfers in a system governed by a WKE with
three-wave resonances. For isotropic systems, in Sec.~\ref{sec:3} we
showed rigorously that the formula encompasses the standard
formula~\eqref{eq:14} for the energy flux as a particular case of
adjacent control volumes in Fourier space ({\it
  cf.}~\eqref{eq:26}). Using the property of {\it detailed energy
  conservation}~\eqref{eq:23}, we showed that the standard flux
formula contains a vanishing part corresponding to self-interactions
({\it cf.}~\eqref{eq:24}).  The new formula always allows us to compute inter-scale energy
  fluxes as integrals of nonzero quantities, also in the stationary
  states -- except, naturally, for equilibrium states for which the
  interaction kernel is vanishing. This provides a general way to
  obtain the {\it Kolmogorov constant} for a three-wave system: In the
  isotropic case it is an alternative route to the KZ
  regularization~\eqref{eq:16} ({\it cf.}  Fig.~\ref{fig:5}); For
  anisotropic systems, this paves the way to the computation of energy
  fluxes, including their prefactors, as shown in Sec.~\ref{sec:6.2}.
      
  We have formalized the theoretical framework that descends from
  formula~\eqref{eq:5} with particular emphasis on the definition of
  the {\it transfer integral} ({\it cf.}~\eqref{eq:29},~\eqref{eq:41}
  and~\eqref{eq:43}). The transfer integral is a decomposition with
  respect to the scale separation of the instantaneous energy
  transfers between a mode and a control volume in Fourier
  space. Using this formalism, we reframed the so-called {\it locality
    conditions} of wave turbulence explicitly in terms of convergence
  conditions of the energy transfer. In Sec.~\ref{sec:7}, we showed
  that the IR and UV convergence conditions for the collision
  integral~\eqref{eq:47}-\eqref{eq:48} are not sufficient to ensure
  convergence of the energy flux. A third condition~\eqref{eq:49} must
  be imposed. For the capillary wave
      turbulence the third condition
  appears to be redundant.

Via the transfer integral, we are able to express the power exchanged
between distant control volumes in Fourier space -- and the scaling of
the power as a function of the scale separation ({\it
  cf.}~\eqref{eq:4.2},\eqref{eq:44}). This is an important
effective metric for the quantification of the locality level
of energy transport \citep{kraichnan1959structure}, which goes beyond
establishing a binary local/nonlocal status of the system. To this
end, we have defined a nondimensional number $w$, the {\it interaction
  width} in Fourier space.  Given a closed set of modes $B$ in Fourier
space, the number $w$ quantifies how far away from set $B$ one has to
move in order to include the $95\%$ fraction of the total power
transferred to $B$ via resonant interactions. Equivalently, $w$
quantifies the distance in Fourier space past which the farthest
(i.e. most scale-separated) $5\%$ of the contribution to the energy
transfer to $B$ is confined. The width $w$ is thus defined naturally
in terms of a definite integral of the transfer integral. The $5\%$
threshold is chosen arbitrarily as a means to roughly establish a
negligibility threshold. Moreover, to emphasize the meaning of scale
separation of the energy transfer, we defined $w$ in logarithmic
scale. As an example, consider $B$ as the set of all the modes larger
than a value $k$. A value $w=1$ would mean that $95\%$ of the energy
transferred to $B$ from modes smaller than $k$ comes from the interval $[k/10,k]$, and $5\%$ comes
from $[0,k/10]$. According to our definition, the ``width'' of the
energy transfer would then be of one order of magnitude, or a factor
of $10$.

We have also shown that the link between $w$ and the standard notion
of locality and non-locality is quite direct: for a {\it local}
spectrum $w$ is finite, whereas for a nonlocal spectrum $w$ is
infinite. For {\it local} spectra, the value of $w$ gives an indication on the
range of scales that is necessary if one hopes to observe a wave
turbulence cascade.  This opens the possibility to estimating
theoretically the width of the transition region between the inertial
range and the dissipation and forcing regions ({\it cf.}
Secs.~\ref{sec:4.2},~\ref{sec:6.2}), improving the current
understanding of the realizability conditions of KZ spectra. We
believe quantifications in this vein to be relevant for experiments of
wave turbulence, where the range of available scales is limited and it
is important to evaluate whether the scale separation between the
forcing and the dissipation regions is sufficiently large for the
onset of an in-between inertial range \citep{deike2014energy,hassaini2019elastic,monsalve2020quantitative,davis2020succession,rodda2022experimental}.  In the examples considered in
this manuscript, we have obtained values of $w$ of about $0.3$ near
equilibrium and $0.7$ at the KZ solution for the surface capillary
waves. For internal waves, there are two directions for energy
transfers: horizontally, we have $w\simeq2.1$ for the stationary
solution and $w\to\infty$ for the Garrett and Munk scale invariant
limit (logarithmic divergence); vertically, we have $w\simeq0.7$ for
the stationary solution and $w\simeq0.5$ for the Garrett and Munk
spectrum of internal waves.

For the isotropic example of surface capillary waves, where there are
about $2$ orders of magnitude of total available frequencies, our
results imply that for the KZ solution an independent inertial range
should have about a factor of $5$ of separation between both the
forcing and the dissipation regions. Therefore, one is likely to
observe a proper wave turbulence solution associated to the KZ
solution over a window of less than a decade of width in the frequency
domain.

Then, the more complex anisotropic example of oceanic internal waves was chosen to show the potential of applicability of our method based on the {\it Main Statement}~\eqref{eq:5}. In particular, the method expands our capability to calculate energy fluxes in several ways: (i) For stationary solutions that differ from the KZ solution (as shown for the solution $a=3.69,b=0$); (ii) for non-stationary transient solutions (as shown for the scale-invariant regime of the  Garrett and Munk spectrum); (iii) for solutions that are mathematically {\it nonlocal}, but after regularization by a physical cutoff are associated with a finite energy flux that is important to quantify (albeit with strong dependence on the cutoff itself; this was also shown for the Garrett and Munk spectrum); (iv) the method applies also to systems that do not satisfy scale invariance.

In summary:
\begin{itemize}
\item We have calculated the amount of energy exchanged between two
  disjoint sets of wavenumbers in Fourier space. This amount is given by the
  {\it Main Statement} in Sec.~\ref{sec:2} and equation~\eqref{eq:5}.
\item We have rederived the classical formula~\eqref{eq:14} for the flux of
  energy in scale-invariant isotropic systems. The classical formula
  needs to be regularized for the KZ stationary state by de L'H\^opital's rule, as it has a
  $0/0$ indeterminacy.
\item Our formalism applies to a more general case:
  non-scale-invariant, not isotropic, not necessarily stationary
  cases. The formula for the energy fluxes does not
      need to be regularized as it is a well defined integral of a nonzero quantity.
\item Our formalism allows us to characterize the level of locality of
  a system, by use of what we defined as the {\it transfer integral}.
\item We therefore introduced the number $w$,
  Eq.~\eqref{eq:7.7}, which characterizes how many orders of magnitude
  of separation in Fourier space are necessary for two sets of modes
  to not be directly communicating.
\item The values of $w$ calculated in this manuscript show that a fair
  amount of ``teleportation'' in Fourier space is present also in
  applications where the transport is traditionally considered fully
  local. We believe that the estimate of $w$ is
    important for the
  interpretation of wave turbulence experiments.
\item The example of surface capillary waves was used to illustrate the
application of the {\it Main Statement}~\eqref{eq:5} and the
formalism of the {\it transfer integral} to a well known wave-turbulence problem.
\item We have shown how the {\it transfer integral} relates to the
  Kolmogorov constants of wave turbulence ({\it
    c.f. Eq.~\eqref{eq:inversion}}).  This equation reveals the
  ``inter-scale structure'' of the Kolmogorov constant.
\item We have applied our formalism to the anisotropic problem of the internal waves in the ocean.
\end{itemize}
\smallskip

 In conclusion, the formalism presented here allows quantification
  of instantaneous energy fluxes for wave turbulence systems dominated by
    three-wave resonant interactions. Our formalism does not
    require  stationarity, scale
  invariance, or strict fulfillment of the locality
  conditions.
The possible applications of our formalism include the improvement
    and the development of a first-principles understanding of many
    geophysical wave systems, with far reaching implications for
    weather and climate prediction.

\medskip

{\bf Declaration of Interests.} The authors report no conflict of interest.

    \acknowledgments {{\bf Acknowledgments.}} This research was
    supported by the NSF DMS award 2009418 and by the ONR grant
    N00014-17-1-2852. Discussions with Dr. Kurt Polzin and Nick Salvatore are gratefully
    acknowledged.  We are thankful to Sergey Nazarenko and two
    anonymous reviewers for their insightful comments, which helped us
    to improve significantly the clarity of the manuscript.
  
\bibliographystyle{jfm}%{unsrt}%
\bibliography{references}

%\pagebreak

\appendix
\section{}\label{app:A}

\subsection{Proof of the {\it Detailed energy conservation} property~\eqref{eq:DC}}

Using the definition in~\eqref{eq:1}, we have
 \begin{equation}\label{eq:4a}
\begin{aligned}
{\cal Z}(\bp_{a},\bp_b,\bp_c)&=
  \omega_a (\mR^a_{bc} - \mR^b_{ca} - \mR^c_{ab}) + \omega_b (\mR^b_{ca} - \mR^c_{ab} - \mR^a_{bc}) + \omega_c (\mR^c_{ab} - \mR^a_{bc} - \mR^b_{ca}) \\
 & =(\omega_a-\omega_b-\omega_c) \mR^a_{bc} + (\omega_b-\omega_c-\omega_a) \mR^b_{ca} + (\omega_c-\omega_a-\omega_b) \mR^c_{ab} \,.
\end{aligned}
    \end{equation}
 Since $ \mR^a_{bc}$ contains a $\delta(\omega_a-\omega_b-\omega_c)$, $ \mR^b_{ca}$ contains a $\delta(\omega_b-\omega_c-\omega_a)$ and $ \mR^c_{ab}$ contains a $\delta(\omega_c-\omega_a-\omega_b)$, each of the three terms vanishes identically (indeed, in the sense of distributions), proving Eq.~\eqref{eq:DC}.
 $\qquad \square$.

\subsection{Proof of the {\it Main Statement}~\eqref{eq:5}}
\begin{figure}
\begin{centering}
\includegraphics[width=\linewidth]{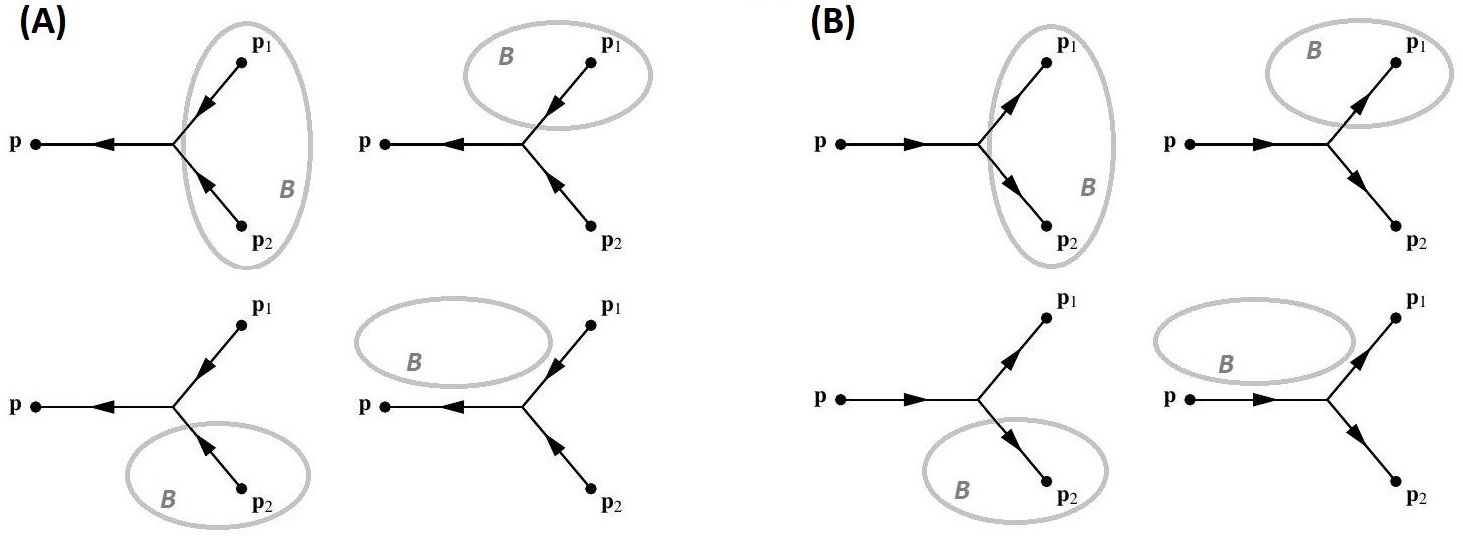}
  \caption{Diagrams associated with the triadic Type I ``sum'' interactions ($\bp=\bp_1+\bp_2$) for a point $\bp\in A$, depending on the sign of the contribution and on whether $\bp_1$ and $\bp_2$ are in set $B$. {\bf(A)} $\mJ^{({\rm I})}(\bp,\bp_1,\bp_2)>0$. {\bf(B)} $\mJ^{({\rm I})}(\bp,\bp_1,\bp_2)<0$.}
  \label{fig:1}
\end{centering}
\end{figure}

In the following,
we provide a proof to the {\it Main Statement}~\eqref{eq:5} in three
steps.

\medskip
\noindent{\bf Step 1}

Consider a triad of type I with two wavenumbers
$\bp_1$ and $\bp_2$ interacting to generate $\bp$, for which Eq.~\eqref{eq:seethelight} holds.
This
relation has a meaning of energy conservation restricted to the particular triad $\bp, \bp_1,\bp_2$ (or {\it detailed energy conservation}). Note that for the action rate of wavenumber $\bp$ we have:
\begin{equation}
	\dot n_\bp|_{012}= \mJ(\bp;\bp_1,\bp_2)=\mR^0_{12}-\mR^1_{02}-\mR^2_{01}=\mR^0_{12}\,,
\end{equation}
since $\mR^1_{02}$ and $\mR^2_{01}$ are vanishing for $\omega_\bp=\omega_1+\omega_2$. For the action rate of wavenumber $\bp_1$ we have:
\begin{equation}
	\dot n_1|_{012}= \mJ(\bp_1;\bp,\bp_2)=\mR^1_{02}-\mR^0_{12}-\mR^2_{01}=-\mR^0_{12}=-\dot n_\bp|_{012}\,.
\end{equation}
Finally, for the action rate of wavenumber $\bp_2$ we have:
\begin{equation}
	\dot n_2|_{012}= \mJ(\bp_2;\bp,\bp_1)=\mR^2_{01}-\mR^0_{12}-\mR^1_{02}=-\mR^0_{12}=-\dot n_\bp|_{012}\,.
\end{equation}
Therefore, $x$ amount of action
with energy $\omega_1x$ interacts with $x$ amount of action with
energy $\omega_2x$ producing $x$ amount of action with energy
$\omega_\bp x = (\omega_1+\omega_2)x$. Equivalently, of the energy supplied to wavenumber $\bp$,
a fraction $\omega_1/\omega_\bp$ comes from wavenumber $\bp_1$ and a fraction $\omega_2/\omega_\bp$ comes from wavenumber $\bp_2$.

\medskip
\noindent{\bf Step 2}

In order to quantify how much of the fraction of the energy being
    transferred to wavenumber $\bp$ through a resonant triad
    $(\bp,\bp_1,\bp_2)$ comes directly from set $B$, we introduce the
    weight function $\chi_{B\rightarrow \bp}^{(l)}(\bp_1,\bp_2)$.
We need to classify all of the possible interactions and define the
weight function $\chi_{B\rightarrow \bp}^{(l)}(\bp_1,\bp_2)$
consistently with detailed energy conservation.

\begin{figure}
\begin{centering}
\includegraphics[width=\linewidth]{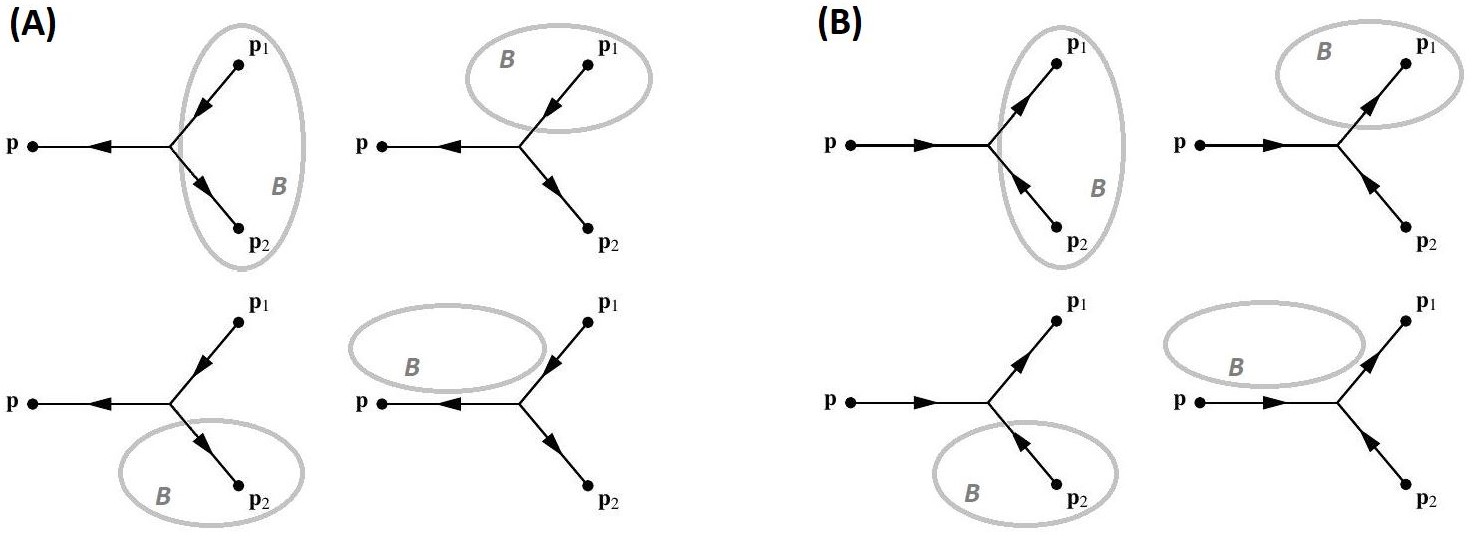}
  \caption{Diagrams associated with the triadic Type II ``difference'' interactions ($\bp=\bp_1-\bp_2$) for a point $\bp\in A$, depending on the sign of the contribution and on whether $\bp_1$ and $\bp_2$ are in set $B$ or not. Left: $\mJ^{({\rm II})}(\bp,\bp_1,\bp_2)>0$. Right: $\mJ^{({\rm II})}(\bp,\bp_1,\bp_2)<0$}
 \label{fig:2}
\end{centering}
\end{figure}

\begin{enumerate}
\item Type I ($\bp=\bp_1+\bp_2$)
\begin{enumerate}
\item $\mJ^{({\rm I})}(\bp,\bp_1,\bp_2)>0$

The four possible configurations analyzed
below are depicted in Fig.~\ref{fig:1}{\bf(A)}. The weight quantifies
what fraction of the energy transferred to $\bp$ comes from set $B$.
\begin{enumerate}
\item $\bp_1\in B$, $\bp_2\in B$: all of the energy going to $\bp$
  comes from $B$, and therefore $\chi_{B\rightarrow \bp}^{({\rm
      I})}(\bp_1,\bp_2)=1$;
\item $\bp_1\in B$, $\bp_2\notin B$: of the energy going to $\bp$,
  only the fraction contained in $\bp_1$ comes from $B$, and therefore
  $\chi_{B\rightarrow \bp}^{({\rm
      I})}(\bp_1,\bp_2)=\omega_1/\omega_\bp =
  \omega_1/(\omega_1+\omega_2)$;
\item $\bp_1\notin B$, $\bp_2\in B$: like in the previous case, but
  exchanging the indices $1$ and $2$, therefore $\chi_{B\rightarrow
    \bp}^{({\rm I})}(\bp_1,\bp_2)=\omega_2/\omega_\bp =
  \omega_2/(\omega_1+\omega_2)$;
\item $\bp_1\notin B$, $\bp_2\notin B$: none of the energy going to
  $\bp$ comes from $B$, and therefore $\chi_{B\rightarrow \bp}^{({\rm
    I})}(\bp_1,\bp_2)=0$.
\end{enumerate}
\item $\mJ^{({\rm I})}(\bp,\bp_1,\bp_2)<0$

The four possible configurations analyzed below are depicted in
Fig.~\ref{fig:1}{\bf(B)}. Since the resulting contribution
to $n_p$ is negative, the weight quantifies what fraction of
the energy lost from wavenumber $\bp$ is transferred to set $B$.
\begin{enumerate}
\item $\bp_1\in B$, $\bp_2\in B$: all of the energy lost from $\bp$ is
  transferred to $B$, and therefore $\chi_{B\rightarrow \bp}^{({\rm
      I})}(\bp_1,\bp_2)=1$;
\item $\bp_1\in B$, $\bp_2\notin B$: of the energy lost from $\bp$,
  only the fraction contained in $\bp_1$ is transferred to $B$, and
  therefore $\chi_{B\rightarrow \bp}^{({\rm
      I})}(\bp_1,\bp_2)=\omega_1/\omega_\bp =
  \omega_1/(\omega_1+\omega_2)$;
\item $\bp_1\notin B$, $\bp_2\in B$: like in the previous case, but
  exchanging the indices $1$ and $2$, therefore $\chi_{B\rightarrow
    \bp}^{({\rm I)}}(\bp_1,\bp_2)=\omega_2/\omega_\bp =
  \omega_2/(\omega_1+\omega_2)$;
\item $\bp_1\notin B$, $\bp_2\notin B$: none of the energy lost from
  $\bp$ is transferred to $B$, and therefore $\chi_{B\rightarrow
  \bp}^{({\rm I})}(\bp_1,\bp_2)=0$.
\end{enumerate}
\end{enumerate}
Note that in cases (a) and (b) the values taken by the weight in the
four sub-cases (i), (ii), (iii), (iv) are respectively the same,
independent of whether the contribution is positive or negative. These
weights are summarized in Table~\ref{tab:1}.

\item Type II ($\bp=\bp_1-\bp_2$)
\begin{enumerate}
\item $\mJ^{({\rm II})}(\bp,\bp_1,\bp_2)>0$

The four possible configurations analyzed below are depicted in Fig.~\ref{fig:2}{\bf(A)}. The weight quantifies what fraction of the energy transferred  to $\bp$ comes from set $B$. Notice that wavenumbers $\bp$ and $\bp_2$ are ``generated'' by a decay of wavenumber $\bp_1$, but there is no net energy exchange between $\bp$ and $\bp_2$.
\begin{enumerate}
\item $\bp_1\in B$, $\bp_2\in B$: all of the energy going to $\bp$ originates from point $\bp_1$, which is in set $B$, and therefore $\chi_{B\rightarrow \bp}^{({\rm II})}(\bp_1,\bp_2)=1$;
\item $\bp_1\in B$, $\bp_2\notin B$: like above, again all of the energy that is transferred to $\bp$ originates from point $\bp_1\in B$, and therefore $\chi_{B\rightarrow \bp}^{({\rm II})}(\bp_1,\bp_2)=1$;
\item $\bp_1\notin B$, $\bp_2\in B$: all of the energy that is transferred to $\bp$ originates from point $\bp_1\notin B$, and therefore $\chi_{B\rightarrow \bp}^{({\rm II})}(\bp_1,\bp_2)=0$;
\item $\bp_1\notin B$, $\bp_2\notin B$: none of the energy going to $\bp$ comes from $B$, and therefore $\chi_{B\rightarrow \bp}^{({\rm II})}(\bp_1,\bp_2)=0$.
\end{enumerate}
\item $\mJ^{({\rm II})}(\bp,\bp_1,\bp_2)<0$

The four possible configurations analyzed below are depicted in Fig.~\ref{fig:2}{\bf(B)}. Since the contribution is negative, the weight quantifies what fraction of the energy lost from wavenumber $\bp$ is transferred to set $B$.
Again, notice that wavenumbers $\bp$ and $\bp_2$ interact together to ``generate a wave'' of wavenumber $\bp_1$, but there is no net energy exchange between $\bp$ and $\bp_2$.
\begin{enumerate}
\item $\bp_1\in B$, $\bp_2\in B$: all of the energy lost from $\bp$ ends up being transferred to $\bp_1$, which is in set $B$, and therefore $\chi_{B\rightarrow \bp}^{({\rm II})}(\bp_1,\bp_2)=1$;
\item $\bp_1\in B$, $\bp_2\notin B$: like above, again all of the energy lost from $\bp$ ends up being transferred to $\bp_1\in B$, and therefore $\chi_{B\rightarrow \bp}^{({\rm II})}(\bp_1,\bp_2)=1$;
\item $\bp_1\notin B$, $\bp_2\in B$: all of the energy that is lost from $\bp$ is transferref to $\bp_1\notin B$, and therefore $\chi_{B\rightarrow \bp}^{({\rm II})}(\bp_1,\bp_2)=0$;
\item $\bp_1\notin B$, $\bp_2\notin B$: none of the energy lost from $\bp$ is transferred to $B$, and therefore $\chi_{B\rightarrow \bp}^{({\rm II})}(\bp_1,\bp_2)=0$.
\end{enumerate}
\end{enumerate}
Again, in cases (a) and (b) the values taken by the weight in the four sub-cases (i), (ii), (iii), (iv) are respectively the same, independent of the contribution being positive or negative. In particular, the weight is independent of the location of wavenumber $\bp_2$, as summarized in Table~\ref{tab:1}.

\item Type III ($\bp=\bp_2-\bp_1$)

Upon permutation of the indices $1$ and $2$, the situation is identical to Type II resonances, as summarized in Table~\ref{tab:1}.
\end{enumerate}
In all cases, the weight $\chi_{B\rightarrow \bp}^{(l)}(\bp_1,\bp_2)$
is expressed solely as a function of $\bp_1$ and $\bp_2$, independent
of $\bp$. Thus, the dependence on $\bp$ can be dropped from the
notation, indicating the characteristic interaction weight simply by
$\chi_{B}^{(l)}(\bp_1,\bp_2)$ in Table~\ref{tab:1}.

\medskip
\noindent{\bf Step 3}

Integrating the interaction kernel  multiplied by the weighting function $\chi_B$ over all possible combinations of $\bp_1$ and $\bp_2$, we obtain
the total energy density time increment of wavenumber $\bp$
corresponding to direct outflow of energy from set $B$,
\begin{equation}\label{eq:10b}
	\mP_{B\rightarrow \bp} = \omega_\bp \int_{\bbR^d\times \bbR^d}d\bp_1 d\bp_2
        \sum_l \chi_{B\rightarrow \bp}^{(l)} (\bp_1,\bp_2)
        \mJ^{(l)}(\bp,\bp_1,\bp_2)\,.
\end{equation}
This expression is valid for all $\bp\in A$, for a given closed set
$A$ such that $A\cap B=\emptyset$. Performing an outer integration
over all $\bp\in A$, we obtain the total instantaneous net flow of
spectral energy density per unit time (in short, the power) from $B$
to $A$,
\begin{equation}
	\mP_{B\rightarrow A} = \int_A d\bp \;\mP_{B\rightarrow \bp}\,.
\end{equation}
By energy conservation, this equals the opposite of the power from $A$
to $B$,
\begin{equation}
	\mP_{A\rightarrow B} = - \mP_{B\rightarrow A} = -\int_A d\bp \;\mP_{B\rightarrow \bp} \,,
\end{equation}
Using Eq.~\eqref{eq:10b}, this finally proves Eq.~\eqref{eq:5}.$\qquad \square$

\subsection{Proof of Eq.~\eqref{eq:24} (vanishing self interactions)}

Consider any three given frequency values $\omega_a$, $\omega_b$, $\omega_c$, such that $\omega_c<\omega_b<\omega_a<\omega$, with $\omega_a=\omega_b+\omega_c$. In the double integration~\eqref{eq:24}, $\omega'$ and $\omega_1$ can take the three values $\omega_a$, $\omega_b$ and $\omega_c$ in six different combinations:
\begin{itemize}
	\item $\omega' =\omega_a, \omega_1 =\omega_b$, and $\omega' =\omega_a, \omega_1 =\omega_c$, giving $2\omega_a J(\omega_a;\omega_b,\omega_c)$;
	\item $\omega' =\omega_b, \omega_1 =\omega_a$, and $\omega' =\omega_b, \omega_1 =\omega_c$, giving $2\omega_b J(\omega_b;\omega_a,\omega_c)$;
\item $\omega' =\omega_c, \omega_1 =\omega_a$, and $\omega' =\omega_c, \omega_1 =\omega_b$, giving $2\omega_c J(\omega_c;\omega_a,\omega_b)$.
\end{itemize}
Thus, the contribution to the integral~\eqref{eq:24} from the triad $\omega_a$, $\omega_b$, $\omega_c$, is given by
\begin{equation}\label{eq:25}
	2\left[\omega_a J(\omega_a;\omega_b,\omega_c) + \omega_b J(\omega_b;\omega_a,\omega_c) + \omega_c J(\omega_c;\omega_a,\omega_b)\right] = 0\,,
\end{equation}
a vanishing contribution by the detailed conservation property~\eqref{eq:23}. Since this is true for any arbitrary choice of $\omega_a,\omega_b,\omega_c$, Eq.~\eqref{eq:24} follows.$\qquad \square$

\section{Detailed conservation for isotropic systems}\label{app:B}
\begin{figure}
\begin{centering}
\includegraphics[width=.55\linewidth]{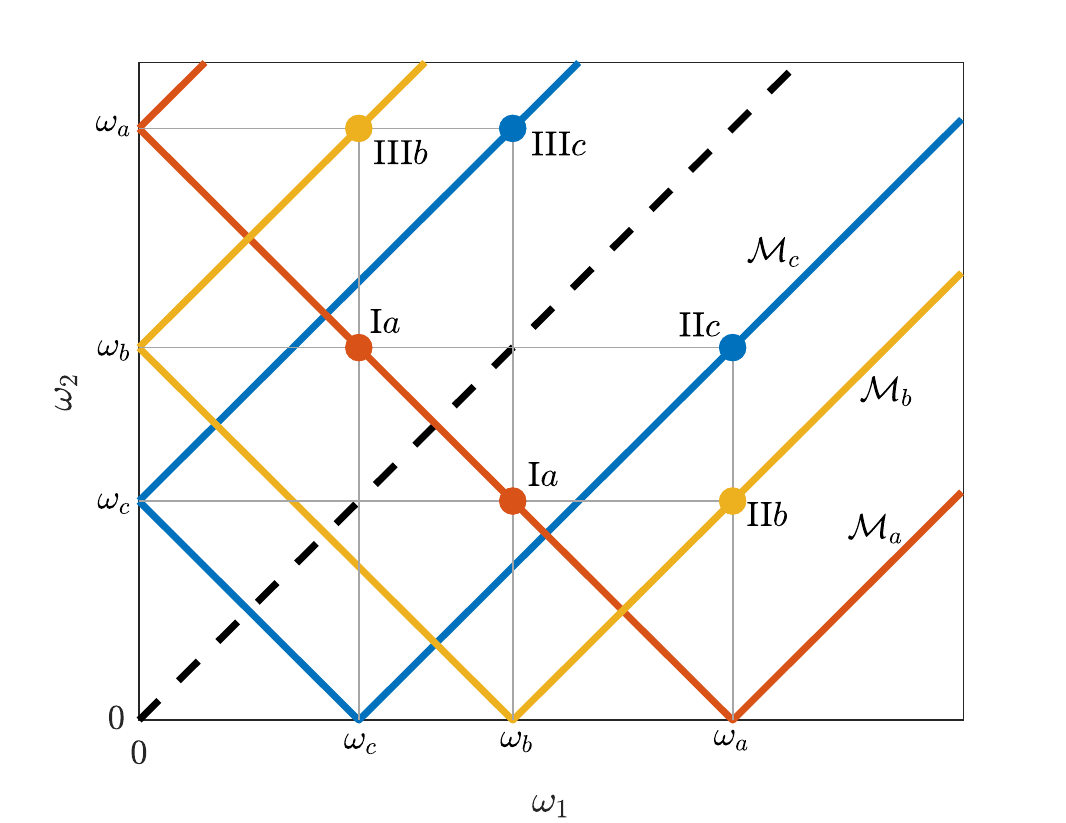}%
  \caption{Representation of the resonant solutions of wavenumbers $\bp_a,\bp_b,\bp_c$, such that $\bp_a=\bp_b+\bp_c$ and $\omega_a=\omega_b+\omega_c$. In the $\omega_1-\omega_2$ space, there are six solutions: two for each of the resonant manifolds corresponding to $\omega=\omega_a$, $\omega=\omega_b$ and $\omega=\omega_c$.}
 \label{fig:4}
\end{centering}
\end{figure}
\begin{quote}
{\bf Property: Detailed conservation for isotropic wave turbulence. }
\it Any triad of wavenumbers $\bp, \bp_1, \bp_2$ on the resonant
    manifold is internally conservative, i.e. it satisfies
\end{quote}
\begin{equation}\label{eq:23a}
\omega_a J(\omega_a;\omega_b,\omega_c) + \omega_b J(\omega_b;\omega_a,\omega_c) + \omega_c J(\omega_c;\omega_a,\omega_b) =0\,.
\end{equation}
{\bf Proof. }  Consider three fixed values of frequency
$\omega_a>\omega_b>\omega_c$ satisfying the resonance condition --
because they are positive, the only possibility is that
$\omega_a=\omega_b+\omega_c$.  Since $\omega_a$ is the largest
frequency, we have $\omega_a J(\omega_a;\omega_b,\omega_c) = \omega_a
J^{({\rm I})}(\omega_a;\omega_b,\omega_c)=\omega_aR^a_{bc}$, using the
definitions in Eq.~\eqref{eq:17}. Graphically, this condition is shown
in Fig.~\ref{fig:4} as the red points on the branch I of the resonant
manifold $\mM_a$ built on $\omega_a$. One has horizontal coordinate
$\omega_1=\omega_b$ and vertical coordinate $\omega_2=\omega_c$, and
the other is its symmetric with respect to the main diagonal. The same
solutions can be represented as the yellow points on the branches II
and III of the resonant manifold $\mM_b$ built on $\omega_b$. Because
of symmetry, for each of these points we have $\omega_b
J(\omega_b;\omega_a,\omega_c)=\omega_b J^{({\rm
    II})}(\omega_b;\omega_a,\omega_c)=-\omega_b R^a_{bc}$. Analogous
reasoning allows us to express the contribution from the two resonant
solutions on $\mM_c$ as $\omega_c J(\omega_c;\omega_a,\omega_b)=
\omega_c J^{({\rm III})}(\omega_c;\omega_a,\omega_b)=-\omega_c
R^a_{bc}$.  Notice that all three cases have two independent
solutions, which can be accounted for as the same solution (in the
region $\omega_1>\omega_2$) by reflection along the main diagonal and
multiplication by a factor of $2$.

Putting the above expressions together, we obtain
\begin{equation}
	\begin{aligned}
		& \omega_a J(\omega_a;\omega_b,\omega_c) + \omega_b J(\omega_b;\omega_a,\omega_c) + \omega_c J(\omega_c;\omega_a,\omega_b) \\
		& \qquad\qquad	\qquad \qquad= \left( \omega_a - \omega_b - \omega_c \right) R^a_{b,c} = \left( \omega_a - \omega_a \right) R^a_{b,c} = 0\,,
	\end{aligned}
\end{equation}
which proves detailed conservation for isotropic systems.$\qquad\square$

\section{Ultraviolet and infrared
integrability conditions for
  capillary
    waves \label{app:C}}

Starting from the expression~(27) in~\cite{ZakharovPushkarevPhysicaD}
and assuming a power-law solution
$n(\omega)=\omega^{-x}$, we  write the nondimensional
collision operator of the isotropic (after angle-averaging)
capillary-wave problem as
\begin{equation}\label{eq:C1}
	\mathcal I(x) = \int_0^{1} S^0_{12} f^0_{12}/\Delta_2 d\xi - 2 \int_1^{+\infty} S^1_{02} f^1_{02} /\Delta_2 d\xi\,,
\end{equation}
where
\begin{equation}
\begin{aligned}
	&S^0_{12} =  (\xi (1-\xi))^{4/3}\Big[ \Big( 1+\frac{1-\xi^{4/3}-(1-\xi)^{4/3}}{2\xi^{2/3}(1-\xi)^{2/3}} \Big)(\xi(1-\xi))^{1/3}\\
	&  -\Big( 1-\frac{1+\xi^{4/3}-(1-\xi)^{4/3}}{2\xi^{2/3}} \Big)\frac{\xi^{1/3}}{1-\xi))^{2/3}} - \Big( 1-\frac{1-\xi^{4/3}+(1-\xi)^{4/3}}{2(1-\xi)^{2/3}} \Big)\frac{(1-\xi))^{1/3}}{\xi^{2/3}} \Big]\,,\\
	&S^1_{02} =  (\xi (\xi-1))^{4/3}\Big[ \Big( 1+\frac{-1+\xi^{4/3}-(\xi-1)^{4/3}}{2(\xi-1)^{2/3}} \Big)\frac{\xi^{1/3}}{\xi-1))^{2/3}}\\
	&  -\Big( 1-\frac{1+\xi^{4/3}-(\xi-1)^{4/3}}{2\xi^{2/3}} \Big)\frac{\xi^{1/3}}{\xi-1))^{2/3}} - \Big( 1-\frac{-1+\xi^{4/3}+(\xi-1)^{4/3}}{2\xi^{2/3}(\xi-1)^{2/3}} \Big)(\xi(\xi-1))^{1/3} \Big]\,,
\end{aligned}
\end{equation}
\begin{equation}
	\Delta_2 = \frac{1}{2}\sqrt{4\xi^{4/3} |\xi-1|^{4/3} - (1-\xi^{4/3} - |\xi-1|^{4/3})^2}\,,
\end{equation}
and
\begin{equation}
\begin{aligned}
	&f^0_{12} =  (\xi(1-\xi))^{-x} - (\xi^{-x} + (1-\xi)^{-x}))\\
	&  f^0_{12} =  (\xi-1)^{-x} - \xi^{-x}(1 + (\xi-1)^{-x}))\,.
\end{aligned}
\end{equation}
\subsection*{Ultraviolet condition}
We first consider integrability of~\eqref{eq:C1} as $\xi\to\infty$.
One can check the following asymptotics as $\xi\to\infty$: $S^1_{02}\sim \tfrac{25}{36}\xi^{4/3}$, $\Delta_2\sim \xi^{2/3}$. Moreover, for $x<1$ and $x\simeq1$, we have that $f^1_{02}\sim 3(1-x)\xi^{-x-5/6}$.
Using these results, we obtain:
\begin{equation}
	S^1_{02}f^1_{02}/\Delta_2 \sim \frac{25}{12} \xi^{-x-1/6}\,,\quad \text{as} \quad \xi\to+\infty
\end{equation}
which is integrable at $+\infty$ if $x>5/6$. This determines the value $\gamma_2=-1/6$ that we use in~\eqref{eq:48} in Section~\ref{sec:7}. However, for $x>1$ the correct asymptotic scaling for the spectrum-dependent term is $f^1_{02}\sim x\xi^{-x-1}$, resulting into a different value  $\gamma_2=-1/3$ which has to be used in~\eqref{eq:49}.

\subsection*{Infrared condition}
Let us now consider the limit as $\xi\to1$. We notice that the
integrand enjoys reflection symmetry in the interval $[0,1]$ with
respect to its center $1/2$. Therefore, we can equivalently consider
integration in the interval $[1/2,1]$ multiplying the first integral
in~\eqref{eq:C1} by a factor of $2$. As $\xi\to1^-$, we pose
$t=1-\xi$, and we have as $t\to0^+$:
\begin{equation}
	2S^0_{12}f^0_{12}/\Delta_2 \simeq 2\Big(\frac{25}{36}t^{8/3} -\frac{35}{27} t^{10/3} - \frac{25}{54} t^{11/3}\Big)t^{-x}\Big(x t + \frac12 x(x+1) t^2\Big) t^{-2/3}
\end{equation}

Likewise, as $\xi\to1^+$, we pose $t=\xi-1$, and we have as $t\to0^+$:
\begin{equation}
	-2S^1_{02}f^1_{02}/\Delta_2 \simeq -2\Big(\frac{25}{36}t^{8/3} -\frac{35}{27} t^{10/3} + \frac{25}{54} t^{11/3}\Big)t^{-x}\Big(x t - \frac12 x(x+1) t^2\Big) t^{-2/3}
\end{equation}
Both expressions have to be integrated in the $t\to0^+$ limit. There
are exact cancellations between the two, and the lowest order terms to
not cancel exactly provides the finite-point singularity
\begin{equation}
	2S^0_{12}f^0_{12}/\Delta_2 -2S^1_{02}f^1_{02}/\Delta_2 \simeq \frac{25}{108} x (3x-1) t^{4-x}\,, \quad \text{as}\quad t\to0\,.
\end{equation}
The corresponding infrared integrability condition is $x<5$.
Notice that if we are looking at the scaling of the singularity as $\xi\to1^+$, as it is done in formula~\eqref{eq:45}, there is no cancellation and the leading order is $O(t^{3-x})$, leading to the value of $\gamma_1=3$ to be used in Eq.~\eqref{eq:47}. Because of double integration in the alternative method of Section~\ref{sec:7}, this leads to the same infrared integrability condition $x<5$.	

\section{Limitations of the dimensional approach in anisotropic systems\label{DoesNotWork}}

        Notice that the KZ solution is the particular case for which
        the $k$-component is independent of $k$, i.e. $F_k=F_k(m)$,
        and the $m$-component is independent of $m$,
        i.e. $F_m=F_m(k)$. However, for
        a general stationary solution for which $F_k=F_k(k,m)$,
        $F_m=F_m(k,m)$, Eq.~\eqref{eq:34} is merely stating that the
        divergence of the flux is zero.  Therefore, this
            approach determines the direction of the flux. The
            magnitude of the flux of energy remains undetermined. Expanding on ideas
        from~\cite{dematteis2022origins}, we
        use Eqs.~\eqref{eq:34}-\eqref{eq:35} and stationarity, with
        the same dimensional ansatz in Eq.~\eqref{eq:36}, to find a
        self consistent closure for the energy flux. For any
        stationary solution with power-law determined by $(a,b)$, the
        energy flux inherits the following form
        \citep{dematteis2022origins}:
\begin{equation}\label{eq:38}
	F_k(k,m)= (1-2b)C k^{7-2a}m^{-2b}\,,\quad F_m(k,m)=(2a-7)C k^{6-2a}m^{1-2b}\,,
\end{equation}
for an arbitrary constant $C$. One can check directly that the energy flux~\eqref{eq:38} is divergence free and satisfies the dimensional constraints of Eqs.~\eqref{eq:34}-\eqref{eq:35}. However, the value of $C$ cannot be determined
by~\eqref{eq:34} (notice that the KZ spectrum is the only
case for which~\eqref{eq:38} is singular, i.e. identically zero, and must
be replaced by~\eqref{eq:37} instead).

The above calculation illustrates the need to quantify the energy flux
for stationary spectra that are not a KZ solution such as the stationary solution $a=3.69,b=0$, since the constant $C$ remains to be determined.
\end{document}